%% file: jetue-epjc.tex
\documentclass[pdftex,twocolumn,epjc3]{svjour3}
\pdfoutput=1

\RequirePackage[T1]{fontenc}
\RequirePackage{graphicx}
\RequirePackage{microtype}

\RequirePackage{tikz}
\RequirePackage{url}
\RequirePackage{placeins}
\RequirePackage{capt-of}

\RequirePackage{xspace}
\RequirePackage{xcolor}

\RequirePackage[amssymb]{SIunits}
\RequirePackage{amssymb}
\RequirePackage{amsmath}
\RequirePackage{setspace,relsize}

\RequirePackage{threeparttable}
\RequirePackage{booktabs}
\RequirePackage{multirow}
\RequirePackage[small,nooneline,center]{subfigure}

\RequirePackage{mathptmx} 
\RequirePackage{latexsym}
\RequirePackage{flushend}
\RequirePackage[numbers,sort&compress]{natbib}
\RequirePackage{afterpage}

\usepackage[hyperindex,breaklinks]{hyperref}

\input{preamble.tex}

\usepackage{todonotes}

\newcommand{\resultsimg}[3][0.45]{\subfigure[]{\includegraphics[width=#1\textwidth]{#2}\label{#3}}}

\makeatletter
\g@addto@macro\bfseries{\boldmath}
\makeatother

\journalname{Eur.~Phys.~J.~C}

\usepackage{preprintcover}
\PreprintCoverPaperTitle{Measurement of the underlying event in jet events\\ from 7\;\TeV proton--proton collisions with the ATLAS detector}
\PreprintIdNumber{CERN-PH-EP-2014-077}
\PreprintJournalName{Eur.~Phys.~J.~C}

\PreprintCoverAbstract{%
  Distributions sensitive to the underlying event in QCD jet events have been
  measured with the ATLAS detector at the LHC, based on $37\;\invpb$ of
  proton--proton collision data collected at a centre-of-mass energy of 7\;\TeV.
  Charged-particle mean \pt and densities of all-particle \et and
  charged-particle multiplicity and \pt have been measured in regions
  azimuthally transverse to the hardest jet in each event.  These are presented
  both as one-dimensional distributions and with their mean values as functions
  of the leading-jet transverse momentum from 20\;\GeV to 800\;\GeV.  The
  correlation of charged-particle mean \pt with charged-particle multiplicity is
  also studied, and the \et densities include the forward rapidity region; these
  features provide extra data constraints for Monte~Carlo modelling of colour
  reconnection and beam-remnant effects respectively.  For the first time,
  underlying event observables have been computed separately for inclusive jet
  and exclusive dijet event selections, allowing more detailed study of the
  interplay of multiple partonic scattering and QCD radiation contributions to
  the underlying event. Comparisons to the predictions of different Monte~Carlo
  models show a need for further model tuning, but the standard approach is
  found to generally reproduce the features of the underlying event in both
  types of event selection.}

\begin{document}

\title{Measurement of the underlying event in jet events\\ from 7\;\TeV proton--proton collisions with the ATLAS detector}
\titlerunning{ATLAS underlying event in 7\;\TeV jet events}

\author{The ATLAS Collaboration}

\institute{CERN, CH-1211 Geneva 23. Switzerland}

\date{Received: DATE / Accepted: DATE}

\maketitle

\begin{abstract}

  Distributions sensitive to the underlying event in QCD jet events have been
  measured with the ATLAS detector at the LHC, based on $37\;\invpb$ of
  proton--proton collision data collected at a centre-of-mass energy of 7\;\TeV.
  Charged-particle mean \pt and densities of all-particle \et and
  charged-particle multiplicity and \pt have been measured in regions
  azimuthally transverse to the hardest jet in each event.  These are presented
  both as one-dimensional distributions and with their mean values as functions
  of the leading-jet transverse momentum from 20\;\GeV to 800\;\GeV.  The
  correlation of charged-particle mean \pt with charged-particle multiplicity is
  also studied, and the \et densities include the forward rapidity region; these
  features provide extra data constraints for Monte~Carlo modelling of colour
  reconnection and beam-remnant effects respectively.  For the first time,
  underlying event observables have been computed separately for inclusive jet
  and exclusive dijet event selections, allowing more detailed study of the
  interplay of multiple partonic scattering and QCD radiation contributions to
  the underlying event. Comparisons to the predictions of different Monte~Carlo
  models show a need for further model tuning, but the standard approach is
  found to generally reproduce the features of the underlying event in both
  types of event selection.

\end{abstract}

\section{Introduction}
\label{sec:intro}

To perform precise measurements or search for new physics phenomena at hadron
colliders it is essential to have a good understanding not only of the hard
scattering process, but also of the accompanying interactions of the rest of the
proton. The aspects of a given collider event not identified with the hard
process are collectively termed the ``underlying event'' (UE). The UE can
receive contributions not only from additional partonic scatters in the same
proton--proton collision (multiple parton interactions, or MPI) and from
quantum chromodynamics (QCD)
colour connections between partons and beam remnants, but also from processes
typically associated with the hard process, such as QCD initial- and final-state
radiation (ISR, FSR). It is impossible, even in principle, to unambiguously
separate the UE from the hard scattering process on an event-by-event basis.
However, observables can be measured which are particularly sensitive to its properties.

The low-momentum QCD processes which dominate the UE are the main type of
interaction in proton--proton collisions. The behaviour of such soft
interactions cannot be calculated reliably with perturbative  QCD methods,
due to the divergence of the QCD coupling at low
scales; it is therefore typically modelled in a phenomenological manner by
Monte~Carlo (MC) event generator programs. These models invariably contain
parameters whose values are not \textit{a priori} known or calculable and must
instead be fitted (``tuned'') to experimental data. Predictivity of UE modelling
is important since the UE forms an irreducible background of particle activity in
all hard-scale processes studied at the LHC. Most new-physics searches operate
in event types with high momentum transfer where the UE has not yet been
measured, hence extension of the measurements into these regions is important,
as is testing the reliability of the UE modelling between hard process types with
different QCD colour flows.

UE observables were previously measured by both the Tevatron and LHC
experiments. The CDF experiment measured the UE with inclusive jet and
Drell--Yan events in Tevatron proton--antiproton ($p\bar{p}$) collision data, at
centre-of-mass energies of $\sqrt{s} = 1.8\;\TeV$\,\cite{Acosta:2004wqa} and
1.96\;\TeV\,\cite{Aaltonen:2010rm}.  The
ATLAS\,\cite{Aad:2010fh,Aad:2011qe,:2012fs,:2012dr}, ALICE\,\cite{ALICE:2011ac}
and CMS\,\cite{Khachatryan:2010pv, Collaboration:2012tb} experiments at the LHC have
thus far measured observables sensitive to the UE in proton--proton ($pp$)
collision data at $\sqrt{s} = 900\;\GeV$ and 7\;\TeV, using track-jets,
leptonically decaying $Z$ bosons, and the highest-\pt charged particle to define
the hard scattering direction in the event.
This paper reports a measurement of UE observables in inclusive jet and
exclusive dijet events, recorded by the ATLAS detector\,\cite{:2008zzm} at the
LHC using proton--proton collisions at a centre-of-mass energy of 7\;\TeV, from
the LHC 2010 run. This study extends the phase-space coverage of previous
studies of the UE by probing a much higher jet-\pT scale (up to $800\;\GeV$),
studying a subset of events with an exclusive dijet topology, and by measuring
the sum of the transverse energy in the full range of the ATLAS calorimeter
acceptance.

This paper is organised according to the following structure. The ATLAS detector
is described in \SecRef{sec:atlasdet}. The observables sensitive to the
underlying event are defined in \SecRef{sec:obs}.  In \SecRef{sec:mcmodels}, the
QCD MC models used in this analysis are discussed.
\SecsRef{sec:selection}--\ref{sec:syst} respectively describe the event
selection, correction of the data back to particle level, and estimation of the
systematic uncertainties.  The results are discussed in \SecRef{sec:res} and
finally the conclusions are presented in \SecRef{sec:summ}.

\section{The ATLAS detector}
\label{sec:atlasdet}

The ATLAS detector is described in detail in \Ref{:2008zzm}. In this
analysis, the trigger system, the tracking detectors, and the calorimeters are of
particular relevance.

The ATLAS inner detector is immersed in the 2\,T axial magnetic field of a superconducting
solenoid, and measures the trajectories of charged particles in the
pseudorapidity range $|\eta| < 2.5$ with full azimuthal
coverage.\footnote{ATLAS uses a right-handed coordinate system with its origin
  at the nominal interaction point (IP) in the centre of the detector and the
  $z$-axis along the beam pipe. The $x$-axis points from the IP to the centre of
  the LHC ring, and the $y$-axis points upward. Cylindrical coordinates
  $(r,\phi)$ are used in the transverse plane, $\phi$ being the azimuthal angle
  around the beam pipe. The pseudorapidity is defined in terms of the polar
  angle $\theta$ as $\eta=-\ln\tan(\theta/2)$.}
The inner detector consists of a silicon pixel detector (pixel), a silicon microstrip detector (SCT)
and a straw-tube transition radiation tracker (TRT). 
The inner detector barrel (endcap) detectors consist of 3 ($2 \times 3$) pixel
layers, 4 ($2 \times 9$) layers of double-sided silicon strip modules, and 73
($2 \times 160$) layers of TRT straws. The pixel, SCT and TRT have $r$--$\phi$
position resolutions of \unit{10}{\micro\metre}, \unit{17}{\micro\metre}, and
\unit{130}{\micro\metre} respectively, and the pixel and SCT have $z$-coordinate
($r$-coordinate) resolutions of \unit{115}{\micro\metre} and
\unit{580}{\micro\metre} respectively in the barrel (endcaps). A track
traversing the barrel typically has 11 silicon hits (3 pixel clusters, and 8
strip clusters), and more than 30 straw tube hits.

Electromagnetic (EM) calorimetry is provided by liquid-argon (LAr) calorimeters.
The barrel ($|\eta|< 1.475$) and endcap ($1.375 < |\eta| < 3.2$) sections have lead
absorbers, and the forward ($3.1 < |\eta| < 4.9$) section (FCal) contains LAr/Cu modules.
The hadronic
calorimeter is divided into four sections:
the barrel ($|\eta|< 0.8$) and
extended barrel ($0.8 < |\eta| < 1.7$), both of which are scintillator/steel
sampling calorimeters; the hadronic endcap ($1.5 < |\eta| < 3.2$), which has
LAr/Cu modules; and the hadronic FCal,
which has LAr/W modules and covers the same $\eta$ range as the EM FCal.
The total calorimeter coverage is $|\eta| <4.9$.

The EM calorimeter has three longitudinal layers (called strip, middle and back)
and a fine segmentation in the lateral direction of the showers within the inner
detector coverage. At high energy, most of the EM shower energy is collected in
the middle layer which has a \etaphi granularity of $\eta \times \phi =
0.025 \times 0.025$. The \etaphi granularity in the hadronic endcap
ranges from $\eta \times \phi = 0.1 \times 0.1$ to $0.2 \times 0.2$. In the
forward calorimeter, the cells are not arranged in projective towers but are
instead aligned parallel to the beam axis. As such the readout granularity is
not constant in \etaphi.

The ATLAS detector has a three-level trigger system: level~1, level~2 and the
event~filter.  Data were taken for this analysis using the single-arm minimum
bias trigger scintillators (MBTS) and central jet-trigger, covering $|\eta|<
3.2$.  The MBTS are mounted at each end of the detector in front of the
liquid-argon endcap-calorimeter cryostats at $z = \pm \unit{3.56}{\metre}$ and
are segmented into eight sectors in azimuth and two rings in pseudorapidity
($2.09 < |\eta| < 2.82$ and $2.82 < |\eta| < 3.84$). Events were triggered by
the MBTS system if at least one hit from either side of the detector was
recorded above threshold.

\section{The underlying event observables}
\label{sec:obs}

The UE observables presented in this paper are constructed separately from
charged-particle tracks with $\pt > 0.5\;\GeV$ and from three-dimensional
clusters of calorimeter cells\,\cite{Lampl:1099735}.

Tracks are required to be within $|\eta| < 2.5$ for all observables
by the acceptance of the ATLAS tracker, while clusters are
constructed separately for pseudorapidity acceptances of $|\eta| < 4.8$ and
$|\eta| < 2.5$ to provide one measurement with full forward coverage and one
compatible with the more restricted acceptance of the tracker.

These detector-level objects were corrected to hadron-level quantities
(i.e. in terms of particles with a mean proper lifetime $\tau > 0.3 \times
10^{-10}\;\text{s}$ either directly produced in the $pp$ interactions or in the
decay of particles with a shorter lifetime) using the definitions given in
\Refs{Aad:2010fh,:2012dr}. The jet correction to hadron level is based on
charged and neutral particles with this lifetime cut, excluding neutrinos. The
selected tracks were corrected to primary charged particles with $\pt >
0.5\;\GeV$ and $|\eta| < 2.5$.  No attempt was made to identify single
calorimeter cell clusters (according to the ATLAS
definition\,\cite{Lampl:1099735}) with primary particles, but momentum sums of
cluster energies were corrected to momentum sums of primary charged particles
with momentum $p > 0.5\;\GeV$ and primary neutral particles (including
neutrinos) with $p > 0.2\;\GeV$.  Lower-momentum particles were not included because
detector simulation indicated that they do not deposit significant energy in the
ATLAS calorimeters,
due to interactions with detector material at smaller radii
and bending in the magnetic
field. Since the
properties of low-momentum particles are not well known or modelled, excluding
them from the hadron-level phase space definition reduces the model dependence
of the correction procedure.

\begin{figure}[t]
  \begin{center}
    \begin{tikzpicture}[>=stealth, very thick, scale=1.1]
      \smaller

      \draw[color=blue!80!black] (0, 0) circle (3.0);
      \draw[rotate= 30, color=gray] (-3.0, 0) -- (3.0, 0);
      \draw[rotate=-30, color=gray] (-3.0, 0) -- (3.0, 0);

      \draw[->, color=black, rotate=-2] (0, 3.5) arc (90:47:3.5) node[right] {$\Delta{\phi}$};
      \draw[->, color=black, rotate=2] (0, 3.5) arc (90:133:3.5) node[left] {$-\Delta{\phi}$};

      \draw[->, color=red, ultra thick] (0, 2) -- (0, 4) node[above] {\textcolor{black}{leading jet}};
      \draw[->, color=green!70!black, ultra thick] (0, -2) -- ( 0.0, -4);
      \draw[->, color=green!70!black, ultra thick, rotate around={-15:(0,-0.3)}] (0, -2) -- ( 0.0, -4);
      \draw[->, color=green!70!black, ultra thick, rotate around={ 15:(0,-0.3)}] (0, -2) -- ( 0.0, -4);

      \draw (0,  1.4) node[text width=2cm] {\begin{center} towards \goodbreak $|\Delta\phi| < 60^\circ$ \end{center}};
      \draw (0, -1.1) node[text width=2cm] {\begin{center} away \goodbreak $|\Delta\phi| > 120^\circ$ \end{center}};
      \draw ( 1.7, 0.3) node[text width=3cm] {\begin{center} transverse \goodbreak $60^\circ < |\Delta\phi| < 120^\circ$ \end{center}};
      \draw (-1.7, 0.3) node[text width=3cm] {\begin{center} transverse \goodbreak $60^\circ < |\Delta\phi| < 120^\circ$ \end{center}};
    \end{tikzpicture}
    \caption{Definition of regions in the azimuthal angle with respect to the
      leading jet. The towards, away and transverse regions are defined in the
      text. The balancing parts of the jet system are indicated with green
      arrows, compatible with the dominant dijet event topology. Multi-jet
      topologies, encountered in the inclusive jet event selection, are expected
      to contribute more substantially to the transverse regions than the
      geometry shown here.}
    \label{fig:ueregions}
  \end{center}
\end{figure}
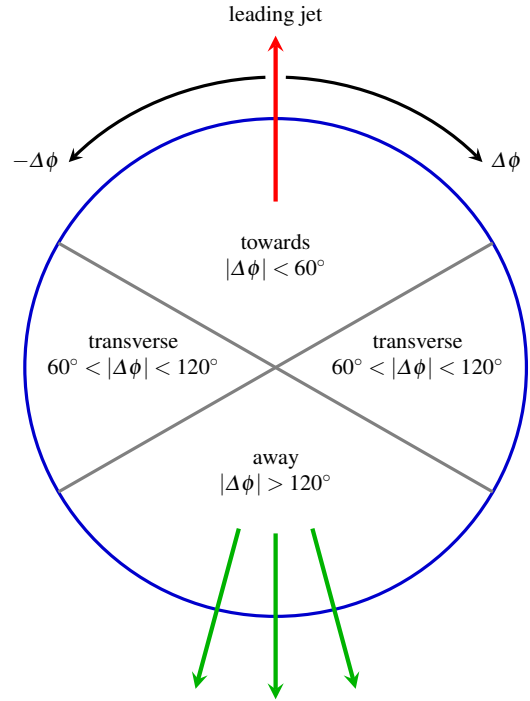

The observables used in this study, defined in \TabRef{tab:obs}, employ the
conventional UE azimuthal division of events into regions relative to the
direction of the ``leading'' object in the event\,\cite{Acosta:2004wqa}.  The
leading object in this case is defined by the 
anti-\kt\,\cite{1126-6708-2008-04-063} jet with distance parameter $R=0.4$ which
has the largest \pt (denoted by \ptlead) after application of the jet selection
criteria described in \SecRef{sec:selection}. At hadron level the jets are
constructed from primary particles as previously defined, excluding
neutrinos. The azimuthal regions are defined with respect to the $\phi$ of the
leading jet: a $120^\circ$ ``towards'' region surrounds the leading jet, an
``away'' region of the same size is azimuthally opposed to it and two
``transverse'' regions each of $60^\circ$ are defined orthogonal to the leading
jet direction. This is illustrated in \FigRef{fig:ueregions}, which displays the
azimuthal distance from the leading jet, $|\dphi|$, used to define the UE
regions.

Since the towards region is dominated by the leading jet and the away region by the
balancing jet (in the dominant dijet configuration), the transverse regions are
the most sensitive to accompanying particle flow, i.e. the UE. In addition, the
transverse regions may be distinguished event-by-event based on which one has
more or less activity, named the ``trans-max'' and ``trans-min'' sides
respectively. The trans-max side is more likely to be affected by wide-angle
emissions associated with the hard process and correspondingly the trans-min
observables have the potential to be more sensitive to soft MPI and beam-remnant
activity.
In this analysis, the trans-min/max definition is specific to the observable
being considered; for example
the trans-max side for the
charged-particle multiplicity (\nch) observable (i.e. the side of the transverse
region which contains more charged particles) can be different from the
trans-max side for the scalar sum of the particle transverse momentum (\ptsum)
observable in the same event.
The difference between trans-max and trans-min sides for a given observable
is referred to as the ``trans-diff'' of that
observable\,\cite{Marchesini:1988hj,Pumplin:1997ix}.  The trans-diff
observables are very sensitive to hard initial- and final-state radiation.

This azimuthal segmentation of events around the leading-jet direction is based
on an assumption of dominant dijet topologies. Measurements of LHC jet
rates\,\cite{Aad:2011tqa} indicate that events with three or more jets
contribute to inclusive jet UE observables; hence there are substantial
contributions to the transverse regions from the hard partonic scattering. While
this subverts the intent of the azimuthally segmented observable definitions,
the resulting interplay of hard and soft event features is itself interesting
and relevant to modelling of hard $pp$ interaction processes at the LHC. Hence,
the transverse-region UE observables are studied both in inclusive jet events where
multi-jet topologies contribute, and in the subset of exclusive dijet events where
higher-order emissions beyond the leading dijet configuration are explicitly
suppressed.

The particle and energy flow observables in the transverse regions are
studied both as one-dimensional distributions, relatively inclusive in the
properties of the hard process, and as ``profile'' histograms which present the
dependence of the mean value of each observable (and its uncertainty) as a
binned function of a hard process property, usually 
\ptlead.

\begin{table*}[bt]
  \caption{Definition of the observables measured in each event at hadron and detector level.
    The hadron-level observables based on momentum use only particles with $\pT > 0.5\;\GeV$,
    and those based on energy use $p > 0.5\;\GeV$ for charged particles and $p > 0.2\;\GeV$
    for neutral particles. All hadron-level definitions are based on particles
    with mean proper lifetime $\tau > 0.3 \times 10^{-10}\;\text{s}$.
    Tracks are selected if they satisfy the criteria described in \protect\SecRef{sec:selection}.
    In the profile plots shown later, these per-event observables are averaged over the events
    in each bin of the profile (i.e. in \ptlead or \nch). The event-ensemble averages shown in
    the profiles are indicated with $\langle \cdot \rangle$ notation, hence the form ``\ptmean''
    is used to distinguish the per-event mean \pt which is profiled as the double mean, $\langle\ptmean\rangle$.
    The trans-max and trans-min
    sub-regions are defined per-event and are also specific to the observable being considered.}

  \begin{center}
    \begin{tabular}{lll}
      \toprule
      Event-wise observable & Particle level & Detector level \\
      \midrule
      \ptlead & \multicolumn{2}{c} {Transverse momentum of the leading jet}  \\
      \addlinespace
      \Nchgdetadphi & \multirow{2}{50mm}{Number of stable charged particles per unit~\etaphi}  & \multirow{2}{50mm}{Number of selected tracks per unit~\etaphi} \\
      \addlinespace
      \addlinespace
      \addlinespace
      \pTsumdetadphi & \multirow{2}{50mm}{Scalar \pt sum of stable charged particles per unit~\etaphi} & \multirow{2}{50mm}{Scalar \pt sum of selected tracks per unit~\etaphi} \\
      \addlinespace
      \addlinespace
      \addlinespace
      mean \pt & \multirow{3}{50mm}{Mean \pt of stable charged particles (at least one charged particle is required)} & \multirow{2}{50mm}{Mean \pt of selected tracks (at least one selected track is required)}  \\
      \addlinespace
      \addlinespace
      \addlinespace
      \addlinespace
      \addlinespace
      \addlinespace
      \ETsumdetadphi & \multirow{2}{50mm}{Scalar \et sum of stable charged and neutral particles per unit~\etaphi} & \multirow{2}{50mm}{Scalar \et sum of selected calorimeter energy clusters per unit~\etaphi} \\
      \addlinespace
      \addlinespace
      \addlinespace
      \addlinespace
      \addlinespace
      \addlinespace
      \bottomrule
    \end{tabular}
    \label{tab:obs}
  \end{center}

\end{table*}


\section{Monte Carlo models of the underlying event}
\label{sec:mcmodels}

In scattering processes modelled by perturbative QCD two-to-two partonic
scatters, at sufficiently low \pt the partonic jet cross-section exceeds that of
the total hadronic cross-section. This apparent problem is resolved by extending
the single-hard-scatter model to include the possibility of multiple partonic
scatters in a given hadron--hadron interaction. In this picture, the ratio of
the partonic jet cross-section to the total cross-section is interpreted as the
mean number of parton interactions in such events. This model is implemented in
several Monte~Carlo event generators, augmented in each implementation by
various phenomenological
extensions\,\cite{Sjostrand:2006za,Sjostrand:2007gs,Gieseke:2011na,Khoze:2010by,Seymour:2013qka}.

This analysis uses simulated inclusive jet events created by the
\pythiasix\,\cite{Sjostrand:2006za}, \pythiaeight\,\cite{Sjostrand:2007gs},
\herwigjimmy\,\cite{Corcella:2002jc}, \herwigpp\,\cite{Bahr:2008pv,
  Gieseke:2011na}, \alpgenherwigjimmy\,\cite{Mangano:2002ea}, and
\powhegpythia\,\cite{Alioli:2010xa} event generators.  All but the last
of these are leading-order parton shower generators, but use different
hadronisation models and different parton shower formalisms; the \pythia family
uses a hadronisation model based on the Lund string and a \pT- or
virtuality-ordered parton shower, while \herwigjimmy and \herwigpp implement a
cluster hadronisation scheme and their parton showers are ordered in emission
angle.\footnote{The \pythia showers include an angular veto on the first ISR
  emission to approximate the colour coherence effect implicit in the \herwig
  angular ordering. The \herwigpp shower additionally uses massive splitting
  functions for heavy-quark emissions.} The \textsc{Fortran}
\herwig\,\cite{Corcella:2002jc} generator by itself does not simulate multiple
partonic interactions; these are added by the \jimmy\,\cite{Butterworth:1996zw}
package. The \alpgen generator provides leading-order multi-leg matrix element
events, i.e. it includes more complex hard process topologies than those used by
the other generators but omits loop-diagram contributions. The \alpgen
partonic events are showered and hadronised by the \herwigjimmy generator
combination, making use of MLM matching\,\cite{Mangano:2002ea} between the
matrix element and parton shower to avoid double-counting of jet production
mechanisms. A related matching process is used to interface \pythiasix to
the next-to-leading-order (NLO) \powheg generator, where the matching scheme
avoids both double-counting and NLO subtraction singularities. The \powheg
matrix element simulates the partonic dijet process at NLO, i.e. including the
loop correction, while the third (real emission) partonic jet is calculated at
leading order.  Revision~r2169 of \powheg is used; this version contains an
important modification to the original \powheg dijet matching which
reduces the incidence of spikes in physical distributions\,\cite{Aad:2013lpa,Nason:2013uba}.

\begin{table*}[ht]
  \singlespacing
  \caption{Details of the MC models used in this paper. It should be noted that
    all tunes use data from different experiments for constraining different
    processes, but for brevity only the data which had most weight in each tune is
    listed. A ``main data'' value of ``LHC'' indicates data taken at $\sqrt{s}=7$\;\TeV,
    although $\sqrt{s}=900$\;\GeV data were also included with much smaller weight in the
    ATLAS tunes. Some tunes are focused on describing the minimum bias (MB) distributions
    better, while the rest are tuned to describe the underlying event (UE) distributions,
    as indicated in ``focus''. The detector-simulated MC configurations used for data
    correction are separated from those used in the results comparison plots, for clarity.
    For the \powhegpythia entry, separate parton distribution functions (PDFs) were used for the matrix element
    and parton shower / multiple scattering aspects of the modelling, indicated
    with ``ME'' and ``PS/MPI'' respectively.}
  \begin{center}
    \resizebox{.98\textwidth}{!}{%
      \begin{tabular}{llllllll}
        \toprule
        Generator & Version & Tune & PDF &  Focus  & Main data  & Used for\\
        \midrule 
        \pythiaeight & 8.157 & AU2~\cite{:mc12py8} & CT10~\cite{Guzzi:2011sv} & UE & LHC & MC/data~comparison \\
        \pythiasix & 6.425 & Perugia\,2011~\cite{Skands:2010ak} & CTEQ5L~\cite{Lai:1999wy}  & UE & LHC  & MC/data~comparison \\
        \pythiasix & 6.421 & DW~\cite{Field:2006gq} & CTEQ5L & UE  &  Tevatron & MC/data~comparison \\
        \herwigpp  &  2.5.1   & UE7-2~\cite{Gieseke:2011xy} & MRST\,\lostst~\cite{Sherstnev:2008dm} & UE & LHC  & MC/data~comparison \\
        \herwigjimmy & 6.510 & AUET2~\cite{:auet} & MRST\,\lostst & UE & LHC & MC/data~comparison \\
        \alpgenherwigjimmy & $2.13+6.510$ & AUET1~\cite{:auet} & CTEQ6L1~\cite{Pumplin:2002vw} & UE & LHC  & MC/data~comparison \\
        \powhegpythia & r$2169+6.425$ & Perugia\,2011 & CT10\,(ME) + CTEQ5L\,(PS/MPI) & UE & LHC  & MC/data~comparison \\
        \midrule
        \pythiasix & 6.425 & AMBT1~\cite{atlastunesto2010data} & MRST\,\lostar~\cite{Sherstnev:2007nd} & MB & Early LHC  & Data~correction \\
        \herwigpp & 2.5.0 & \lostar{}\_\,JETS~\cite{hpptunewiki} & MRST\,\lostar & UE  & Tevatron & Correction~systematics \\
        \bottomrule
      \end{tabular}}
    \label{tab:mcgens}
  \end{center}
\end{table*}

Such models typically introduce several parameters, which are to be tuned to
data at different centre-of-mass energies and for a variety of hadronic
processes. MC tuning in recent years has made extensive use of ATLAS and
Tevatron underlying event and minimum bias data for a variety of event
generators, most notably the \pythia and \herwig families.
In this study we have used tuned MC generators both for data correction and
systematic uncertainties, and for comparison to the detector-corrected data
observables.
The combinations of generators, PDFs and tunes are listed in \TabRef{tab:mcgens}.
No currently available tunes make use of UE data from a $pp$ initial state and
high-\pT jet events as measured here, so the comparisons of corrected data to MC
models give a good indication of the level of predictivity of models tuned to
Tevatron or low-\pT LHC data.

The AMBT1 tune (ATLAS Minimum Bias Tune~1)\,\cite{atlastunesto2010data} of
\pythiasix was the first LHC-data tune constructed by ATLAS, and primarily used
diffraction-suppressed observables from the early ATLAS minimum bias (MB)
measurements\,\cite{Collaboration:2010ir} as input to the tuning.

Perugia\,2011\,\cite{Skands:2010ak} is the latest in the ``Perugia'' set of \pythiasix tunes, which
use early LHC minimum bias and underlying event data and the \pythia \pt-ordered
parton shower. A consistent definition of the strong coupling \alphaS
is used throughout the parton showers for smooth interfacing to multi-leg matrix
element generators, particularly \alpgen. This tune uses
the CTEQ5L1 PDF\,\cite{Lai:1999wy}.

The \pythiasix~DW\,\cite{Field:2006gq} tune uses a virtuality-ordered parton
shower and an MPI model not interleaved with the ISR.  This tune was constructed
to describe CDF Run~II underlying event and Drell-Yan data and also uses the
leading-order CTEQ5L1 PDF; it is included here for comparison to previous
results.

\pythiasix is also used in conjunction with a matched \powheg NLO matrix
element using the CT10 PDF set, as this is the highest perturbative order of QCD
jet event modelling currently available in fully showered/hadronised form. The
Perugia\,2011 tune of \pythiasix is used to shower these \powheg events. While
commonly used together, the Perugia\,2011 tune was constructed
based on leading-order matrix elements only, and without the \powheg matching
modification to the parton shower; the interest in comparing this model to data
is hence to observe whether the \powhegpythia matching procedure has any adverse
effect on UE distributions relative to standalone \pythiasix.

\pythiaeight adds to the established \pythia-family MPI model by interleaving
not only the ISR emission sequence with the MPI scatterings, but also the FSR;
all three processes compete against each other for emission phase space in the
resulting evolution. The AU2~CT10 tune\,\cite{:mc12py8} used here is the variant
of the latest ATLAS \pythiaeight UE tune set intended for use with the NLO CT10
PDF\,\cite{Lai:2010vv}. This configuration is the standard setup for jet event
simulation in current ATLAS production use, with the model tuned to give a very
good description of leading-track UE data at $\sqrt{s}=7\;\TeV$.

Two tunes of \herwigjimmy are used in this study: the ATLAS AUET2~\lostst tune
for standalone \herwigjimmy, and the older AUET1~CTEQ6L1 used in the \alpgen
sample\,\cite{:auet}.  For \herwigpp the standard 7\;\TeV underlying event tune
UE7-2\,\cite{Gieseke:2011xy} for the MRST~\lostst PDF is used. This model
includes a colour reconnection model -- the first for a cluster hadronisation
generator -- and provides a good description of both leading-track UE and
minimum bias data at 7\;\TeV.  The \lostst PDF, like the \lostar PDF used in the
\pythiasix~AMBT1 sample, is one of a series of ``modified leading order'' PDFs
which attempt to mimic full NLO simulated event characteristics
by relaxation of PDF momentum sum rules and use of a $\pT^2$-based
factorisation scale for better compatibility with the approximate resummation
inherent in parton shower algorithms.

The \pythiasix~AMBT1 and \herwigpp\;2.5.0 samples were processed through the ATLAS
detector simulation framework\,\cite{Aad:2010ah}, which is based on
\textsc{Geant4}\,\cite{Agostinelli:2002hh}, then reconstructed and analysed similarly
to the data. These reconstructed events were used to calculate detector
acceptances and efficiencies, to correct the data for detector effects. The
fully simulated and reconstructed samples include overlay of MC pile-up events,
simulated using \pythiaeight with the AM2 tune\,\cite{:mc11}, with the mean number of
interactions distributed according to data conditions.
The \herwigpp sample used for unfolding (i.e. correction of residual detector
effects) was generated using \herwigpp~2.5.0 with the default
\lostar{}\_\,JETS\,\cite{hpptunewiki} tune -- an older configuration than that
shown in the physics comparison plots in this paper. All other MC models
described are used at the generated stable particle level only.


\section{Analysis object selection}
\label{sec:selection}

This analysis uses the full ATLAS 2010 dataset of jet events in proton--proton
collisions at $\sqrt{s} = 7\;\TeV$, in which the average number of multiple $pp$
interactions per bunch crossing (``pile-up'') was much smaller than in
the 2011 data-taking period.
Basic requirements on data quality and the operating conditions of the beam,
the relevant sub-detectors and the triggers
resulted in a dataset corresponding to an
integrated luminosity of $37 \pm 1.3\;\invpb$, uncorrected for trigger prescales\,\cite{Aad:2013ucp}.

To select jet events, two different trigger systems were used: the MBTS trigger
and the central jet-triggers, the latter of which has several \pt
thresholds. The MBTS trigger selects events containing charged particles within
the fiducial acceptance of the minimum bias trigger scintillator, and was used
to select events with jets having transverse momenta in the range
20--60\;\GeV. Jets with \pT greater than 60\;\GeV are above the threshold of
full efficiency of the calorimeter-based jet-trigger, hence this was used to
select jet events above this scale. A single jet-trigger was used for each
\ptlead bin in the observables, specifically the fully efficient ($> 99\%$)
trigger with the smallest possible prescale factor for that
bin\,\cite{Aad:2011fc}.

To reject events due to cosmic-ray muons and other non-collision backgrounds,
events were required to have at least one reconstructed ``primary'' $pp$
interaction vertex, each having at least five tracks satisfying the following
criteria\,\cite{Collaboration:2010ir}:
\begin{itemize}
\item $\pt > 0.5\;\GeV$;
\item $|\eta| < 2.5$;
\item a minimum of one pixel and six SCT
  hits;
\item a hit in the innermost pixel layer, if the corresponding pixel module was
  active;
\item transverse and longitudinal impact parameters with respect to primary
  vertex $|d_0| < 1.5\;\mm$ and $|z_0|\sin \theta < 1.5\;\mm$;
\item for tracks with $\pt > 10\;\GeV$, a $\chi^2$ probability of track fit
  greater than $0.01$ was required in order to remove mismeasured tracks.
\end{itemize}

Events with $\ptlead < 100\;\GeV$ were taken in conditions with negligible
pile-up, while the bulk of events with higher \ptlead were taken in the second
2010 data-taking period, with a mean of 2--3 $pp$ interactions per bunch
crossing.
To reduce the contributions from pile-up, events containing more than one
reconstructed primary vertex with at least two associated tracks were removed.

Finally, to obtain an inclusive jet event selection, each event was required to
contain at least one jet with $\pt > 20\;\GeV$ and rapidity $|y| < 2.8$. The
resulting sample is entirely dominated by pure QCD jet events. For the exclusive
dijet event selection an additional requirement of one and only one subleading
jet was made, where the subleading jet had to pass the same cuts as the leading
jet and also have $\pt > 0.5\,\ptlead$ and $|\dphi| > 2.5$ to the
leading jet.
In both selections the jets were corrected
to account for the calorimeters' response to the deposited energy.

Around 429,000 and 99,000 events were selected for the inclusive jet and
exclusive dijet selections respectively.


\section{Correction to particle level}
\label{sec:corr}

To allow comparison of these results with
theoretical predictions and other experimental
studies, the underlying event distributions need to be corrected for selection
efficiencies and detector resolution effects.  A two-step correction procedure
was used, where first the track efficiency corrections were applied for the
track-based observables, then the remaining detector effects were unfolded to
produce observables at particle level which may be directly compared to MC model
predictions.

\subsection{Track reconstruction efficiency and cluster calibration}
\label{sec:trkeff}

The efficiency
of the ATLAS detector to reconstruct a
charged particle as a track was measured
in  a previous analysis~\cite{Collaboration:2010ir}, with the same track selection as used here.  Each
track was reweighted by the inverse of this efficiency, which also accounts for
the contributions of tracks reconstructed from secondary particles, and
particles whose true kinematics were outside the kinematic range (OKR) of the
selection but which were reconstructed within the acceptance cuts due to
detector resolution. For tracks with $\pT > 500\;\MeV$, the effects of fake
tracks (those constructed from tracker noise and/or hits which were not produced
by a single particle) and OKR migrations were found to be negligible.

For cluster-based observables,
both the hadronic and electromagnetic calorimeter cell clusters were used, with calibrations
as described in \Refs{Cojocaru:2004481} and~\cite{Lampl:1099735}.
The cluster energy was corrected to the momentum of the
charged or neutral hadron, and the simulation was validated using the diphoton
invariant mass distribution, $M_{\gamma\gamma}$, for $\pi^0 \rightarrow \gamma\gamma$ candidates\,\cite{:2012dr}.

\subsection{Unfolding}
\label{sec:unfolding}

Bayesian iterative unfolding\,\cite{D'Agostini:1994zf} was used to correct for
residual detector resolution effects, using the Imagiro~0.9
software package\,\cite{Imagiro}.

The Bayesian iterative unfolding method requires two inputs: a prior probability
distribution for the observable (the MC generator-level distribution is used for
this), and a smearing matrix which relates the measured distribution of an
observable to its hadron-level distribution.  The smearing matrix element $S_{ij}$
is the probability of a particular event from bin~$i$ of the hadron-level
distribution being found in bin~$j$ of the corresponding reconstructed
distribution, as calculated using MC samples. For the profile histogram
observables in this paper, a two-dimensional (2D) histogram was created with dense binning in the
observable whose evolution is being profiled, such that each unfolding bin
corresponds to a 2D range in e.g. \ptlead vs. \ptsum space: the 2D nature of the
bin has no effect on the mathematics of the unfolding procedure. Migrations
involving the $\ptlead < 20\;\GeV$ region, excluded from the analysis phase
space, were handled in detail for the unfolding of observables binned in
\ptlead, but for other observables such in/out migrations were treated as fake
or missing events respectively; this was required because a more detailed approach
would add an extra dimension to the smearing matrix, which would then be too
statistically limited to be usable.

The unfolding process is iterated to avoid dependence on the prior probability
distribution; the corrected data distribution produced in each iteration is used
as the prior for the next.  In this analysis, two iterations were performed since
this gave the smallest residual bias when tested on MC samples while keeping the
statistical uncertainties small.
The central values of the data distributions corrected to hadron level were calculated
using prior distributions and smearing matrices from the \pythiasix~AMBT1 MC sample.

\vspace*{-1\lineskip}


\section{Systematic uncertainties}
\label{sec:syst}

Systematic uncertainties on the measured distributions were assessed using
the same unfolding procedure as for the central values of the observables, but
with the input \ptlead, track \& cluster \pt and track \& cluster weights shifted by
$\pm 1\sigma$ variations in each source of uncertainty.

\begin{table*}[btp]
  \caption{Summary of systematic uncertainties for inclusive jet and exclusive dijet profiles vs. \ptlead.
           The ``efficiency'' uncertainties include
           material uncertainties in the tracker and calorimeter geometry modelling. The ``JES'' uncertainty
           source for jets refers to the jet energy scale calibration procedure\,\cite{Aad:2011he}.}
  \centering
  \begin{tabular}{lllllllll}
    \toprule
    Quantity & \hspace{2ex} & \multicolumn{3}{l}{Inclusive~jets} & \hspace{2ex} & \multicolumn{3}{l}{Exclusive~dijets} \\
    \midrule
    \addlinespace
     & & \multicolumn{3}{l}{Pile-up and merged vertices} & & \multicolumn{3}{l}{Pile-up and merged vertices} \\
    All observables & & \multicolumn{3}{l}{1--3\%} & & \multicolumn{3}{l}{1--5\%} \\
    \midrule
    \addlinespace
    Charged tracks       & & Unfolding & Efficiency & & &  Unfolding & Efficiency & \\
    \ptsum               & & 3\%       & 1--7\%     & & &  3--13\%   & 2--7\%    & \\
    \nch                 & & 1--2\%    & 3--4\%     & & &  3--22\%   & 3--7\%     & \\
    \ptmean              & & 1\%       & 0--4\%     & & &  1--9\%    & 1\%     & \\
    \midrule
    \addlinespace
    Calo clusters        & & Unfolding & Efficiency & & &  Unfolding & Efficiency & \\
    \etsum, $|\eta|<4.8$ & & 2--3\%    & 4--6\%     & & &  5--21\%   & 4--9\%    & \\
    \etsum, $|\eta|<2.5$ & & 3--5\%    & 4--6\%     & & &  1--21\%   & 4--7\%    & \\
    \midrule
    \addlinespace
    Jets                 & & Energy resolution & JES       & Efficiency  & & Energy resolution  & JES      & Efficiency \\
    \ptlead              & & 0.3--1\%          & 0.3--4\%  & 0.1--2\%    & & 0.4--3\%           & 1--3\%   & 0.3--3\%     \\
    \bottomrule
  \end{tabular}
  \label{tab:systematics-vs-leadpt}
\end{table*}

\noindent The following sources of uncertainty are included:
\begin{description}
\item[Jet reconstruction:] These uncertainties -- dependent on the jet
  energy scale calibration procedure, the impact of pile-up on the jet energy, the jet
  reconstruction efficiency and the jet energy resolution -- have been
  calculated as in \Ref{Aad:2011he}. For the UE observables profiled against
  leading-jet \pt, they add up to approximately 1\%.
\item[Track reconstruction-efficiency uncertainty:] Tracking eff\-ici\-ency
  uncertainties were studied in \Ref{Collaboration:2010ir}, the
  two largest being found to be due to the amount of material in the inner detector and the
  consequence of the $\chi^2$ probability cut to remove misreconstructed
  tracks. The effect of uncertainties in the amount of material in the inner detector
  is a $2\%$ relative uncertainty in the efficiency in the barrel region,
  rising to over 7\% for $2.3 < \etamod < 2.5$, for tracks with
  $\pt > 500\;\MeV$.  The maximum difference between the fraction of events in
  data and MC simulation which passed the $\chi^2$ probability cut was found to be
  10\%. This value was taken as a conservative estimate of the systematic
  uncertainty, and was applied to tracks with $\pT > 10\;\GeV$ only.
\item[Cluster reconstruction efficiency:] The accuracy with which the MC samples
  simulate the energy response of the calorimeters to low-energy particles was
  determined separately for electromagnetic and hadronic particles. An average
  is then obtained, using the \pythiasix~AMBT1 prediction of the relative
  contribution to \etsum by different particle types\,\cite{:2012dr}.  For
  electromagnetic particles the systematic uncertainty arises from the
  extraction of the energy scale from fits to the $M_{\gamma\gamma}$
  distributions in $\pi^0 \rightarrow \gamma\gamma$ candidates.  The total
  uncertainty depends on the $|\eta|$ region and is typically 2--4\%, but
  is as large as $15\%$ in the regions where different calorimeter subsystems
  overlap.
  The uncertainty in the energy response for hadronic particles in the
  central region ($|\eta| < 2.4$), where there is good coverage from the inner
  tracking detector, was obtained from studies of
  the ratio of the energy measured by the calorimeter to the track momentum
  measured by the inner detector for
  isolated charged particles and is within a few percent\,\cite{Aad:2011he}.
\item [Pile-up and merged vertices:] The effect of pile-up on the underlying
  event observables after the tight vertex selection was assessed using an MC
  sample with realistic modelling of the pile-up conditions in 2010 data; the
  largest deviation from the sample without pile-up was observed to be 1\%. A
  $\pm 1\%$ shift was hence applied to the values of all observable bins used as
  input to the unfolding, and the resultant shifts on unfolded observables have
  been used as the systematic uncertainty due to merged vertices. This
  uncertainty is less than 2\% for most bins, but increases to approximately 5\%
  at high \ptlead.
\item[Unfolding:] The uncertainty due to model-dependence of the unfolding
  procedure was taken as the difference between the results of unfolding
  with each of the two MC samples, \pythiasix AMBT1 and \herwigpp. The dominant
  effect arises from modelling of low \pt QCD radiation, which produces a
  typical uncertainty of 1\% but can reach 20\% at the edges of the fiducial
  phase space studied here.
  An additional uncertainty was estimated using two different priors: the
  unmodified generator-level distribution, and the generator-level distribution
  after reweighting so that the reconstructed distribution matches the data.
  This uncertainty is less than 1\% throughout the fiducial phase space of the
  measurement.
\end{description}

Systematic uncertainty contributions from electronic noise
and beam-induced background, simulation of the primary vertex position, and
simulation of the trigger selection were also
considered and were found to be negligible.  Table~\ref{tab:systematics-vs-leadpt}
summarises the sizes of contributions to systematic uncertainties for the UE
profile observables with both the inclusive jet and
exclusive dijet selections. The \meanptmean vs. \nch profile observables have flat
symmetric uncertainties of 1\% each for unfolding and reconstruction efficiency,
increasing to $\sim 10\%$ in the first and last bins of \nch.

\FloatBarrier


\begin{figure*}[tbp]
  \centering
  \vspace{0.5cm}
   \resultsimg{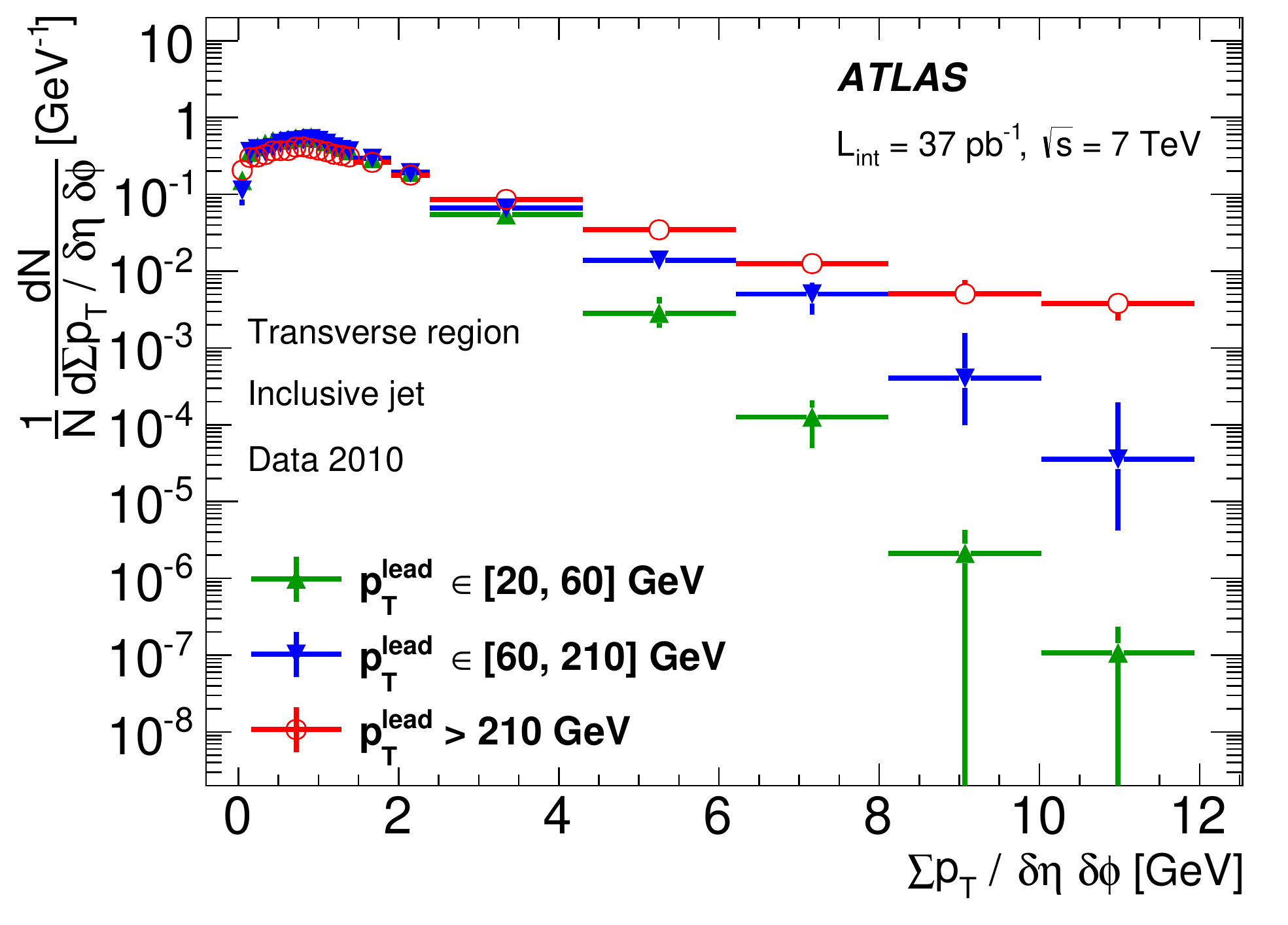}{fig:PtSlices_PtSum}
   \resultsimg{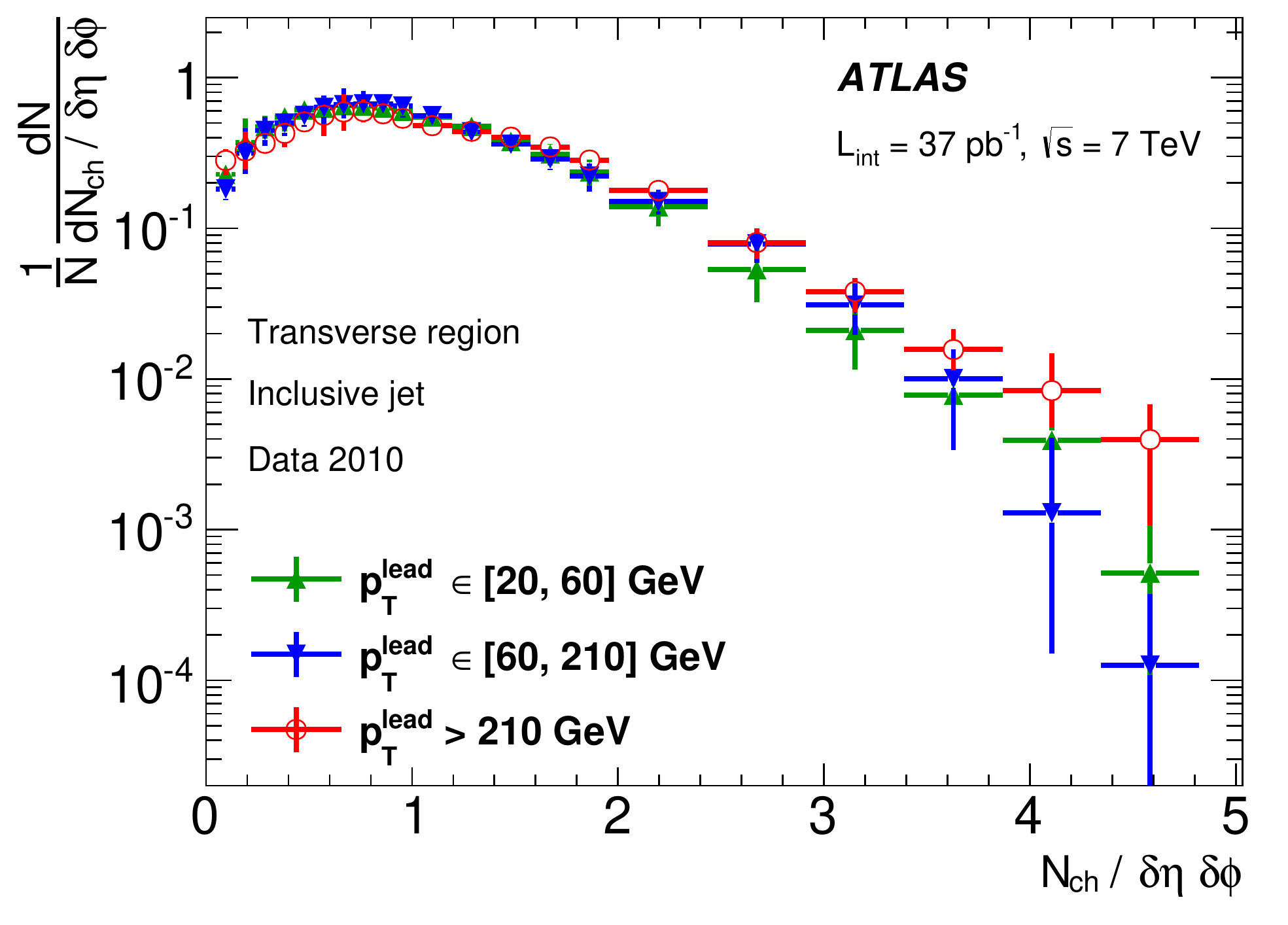}{fig:PtSlices_Nch}\\
   \resultsimg{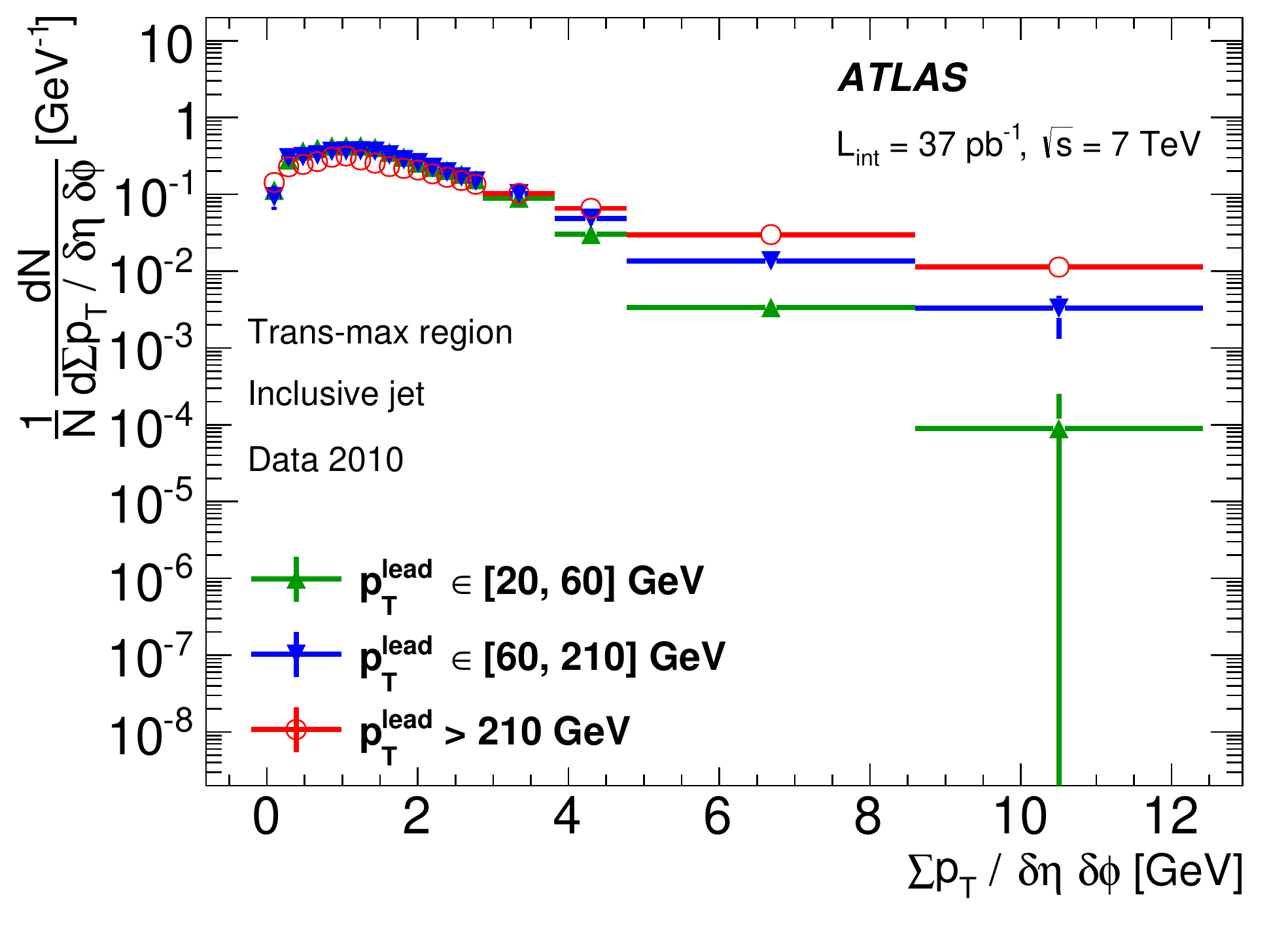}{fig:PtSlices_PtSumMax}
   \resultsimg{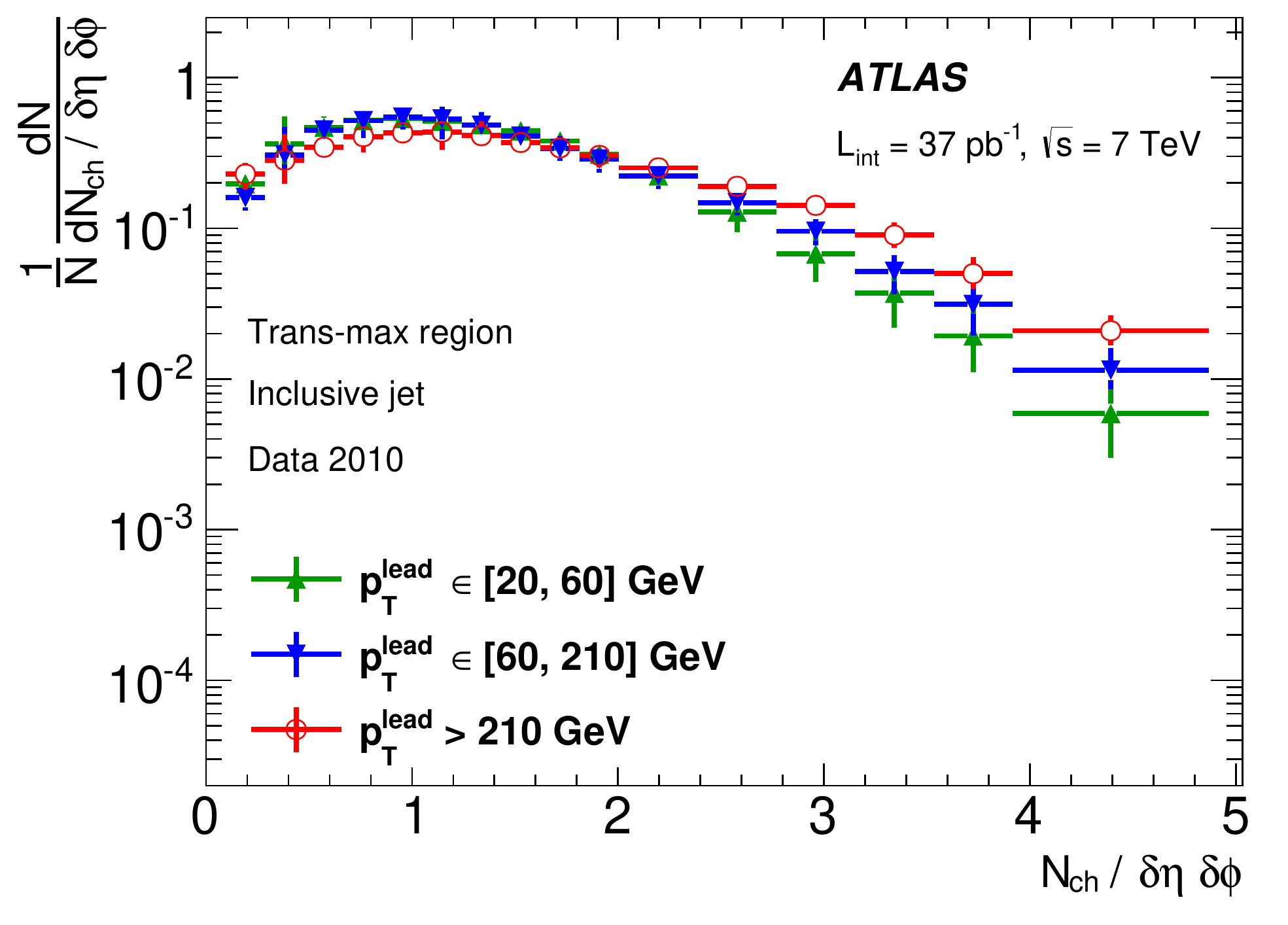}{fig:PtSlices_NchMax}\\
   \resultsimg{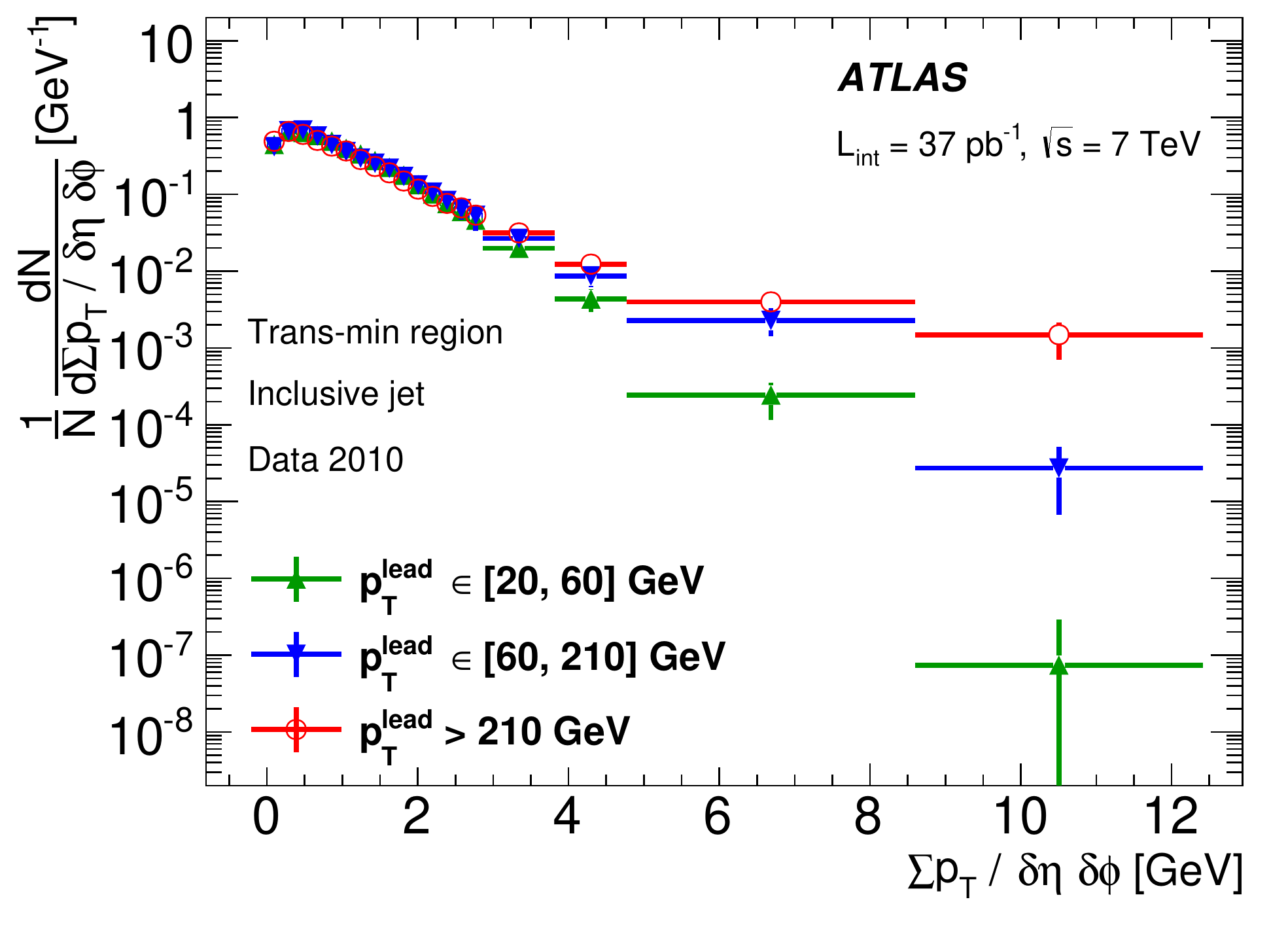}{fig:PtSlices_PtSumMin}
   \resultsimg{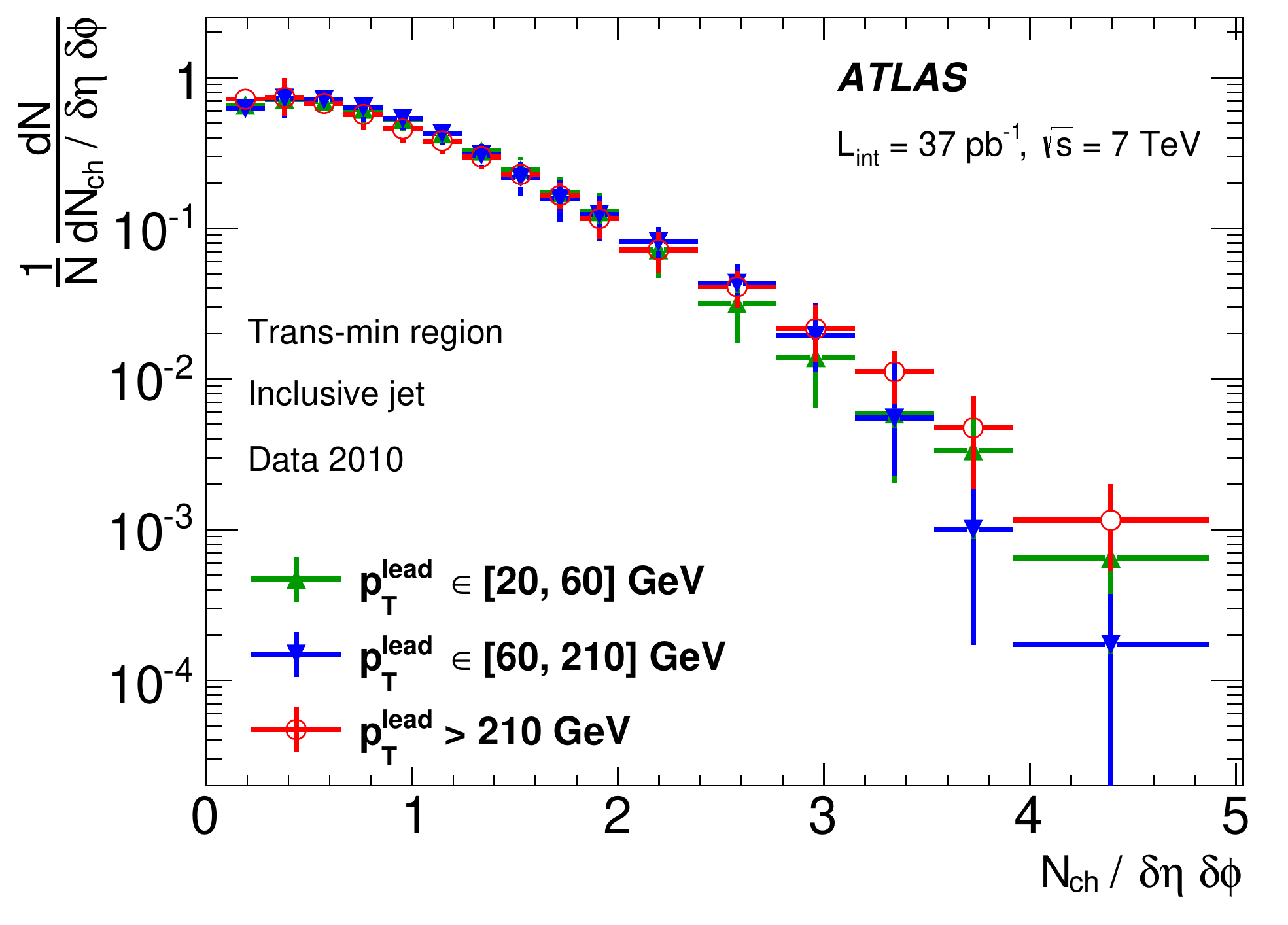}{fig:PtSlices_NchMin}
   \caption{Comparison of \ptsum (left column) and \Nch (right column) density distributions
  for different \ptlead requirements, data only.
  The jets are required to have \protect\pT of at least $20\;\GeV$, and
    be within $|y| < 2.8$, whereas the charged particles have at least a \protect\pT of $0.5\;\GeV$
    and  $|\eta| < 2.5$.
  The top, middle and bottom rows, respectively, show the transverse,
   trans-max and trans-min regions. Each \ptlead slice is independently normalised to unity via division by the number of events in the slice, $N$.
   The error bars show the combined statistical and systematic uncertainty.}
  \label{fig:nchgsumpt-dataonly}
\end{figure*}

\begin{figure*}[tbp]
  \vspace{0.5cm}
  \begin{center}
    \resultsimg{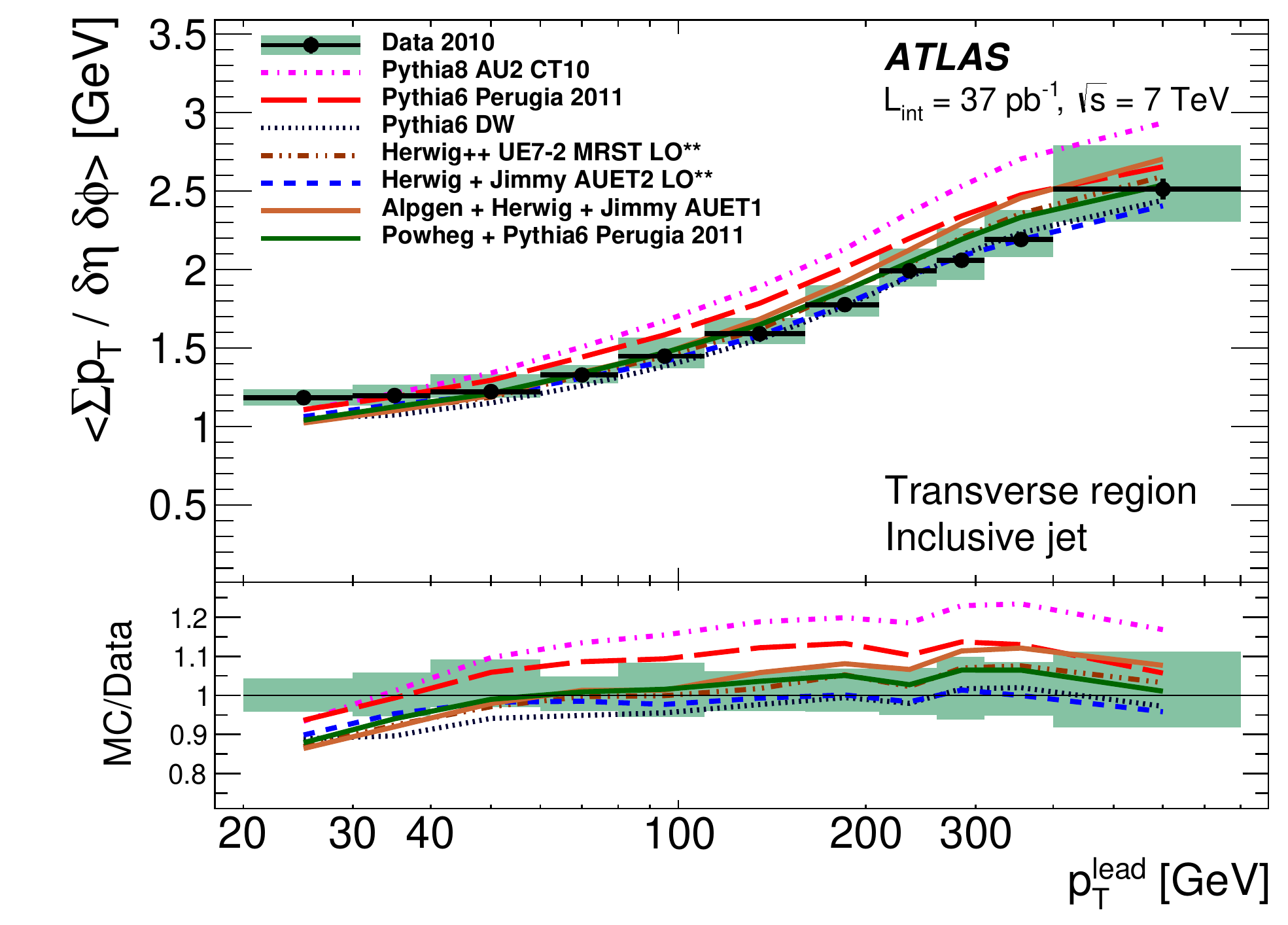}{fig:in-sumpt}
    \resultsimg{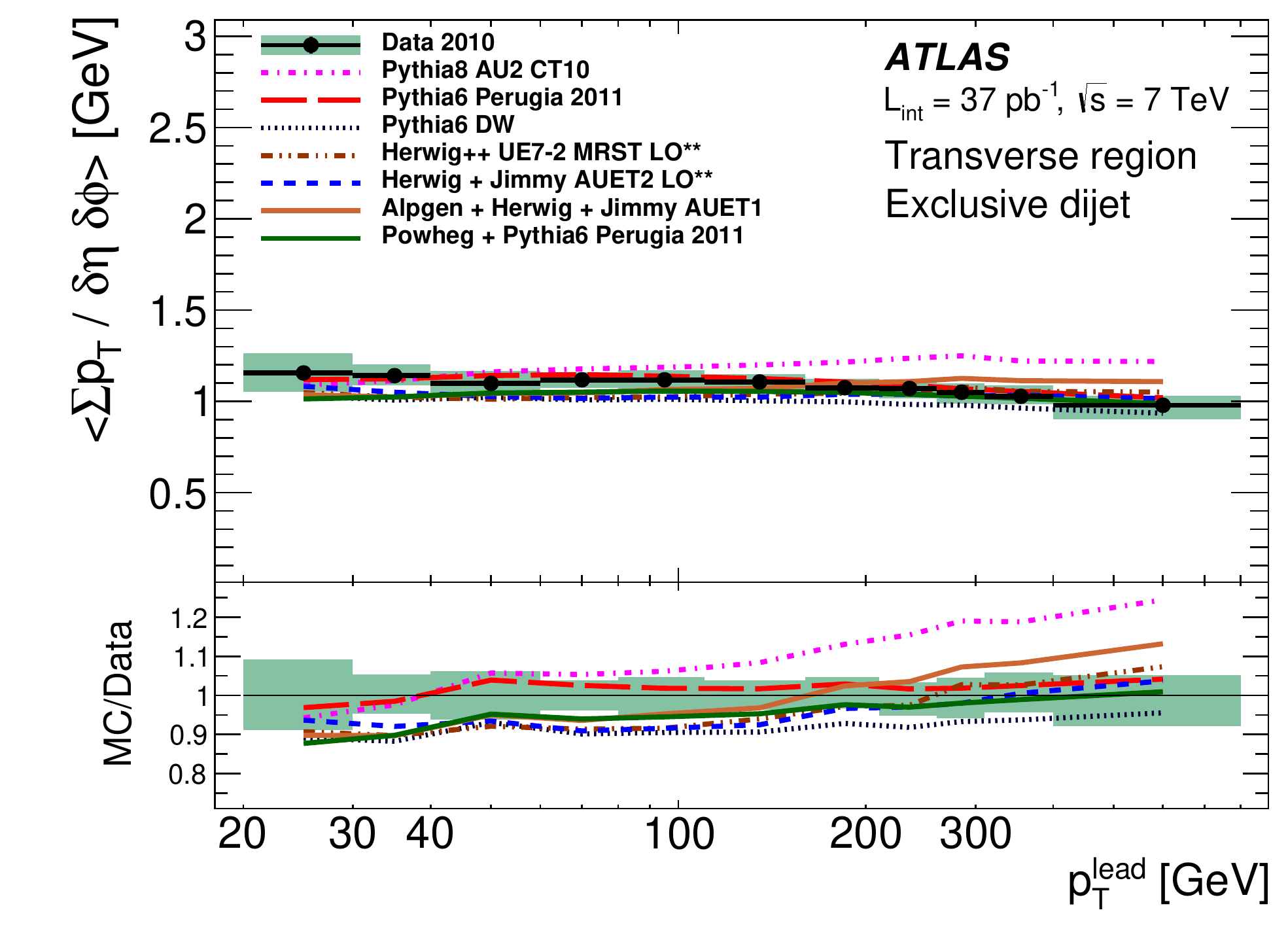}{fig:ex-sumpt}
    \resultsimg{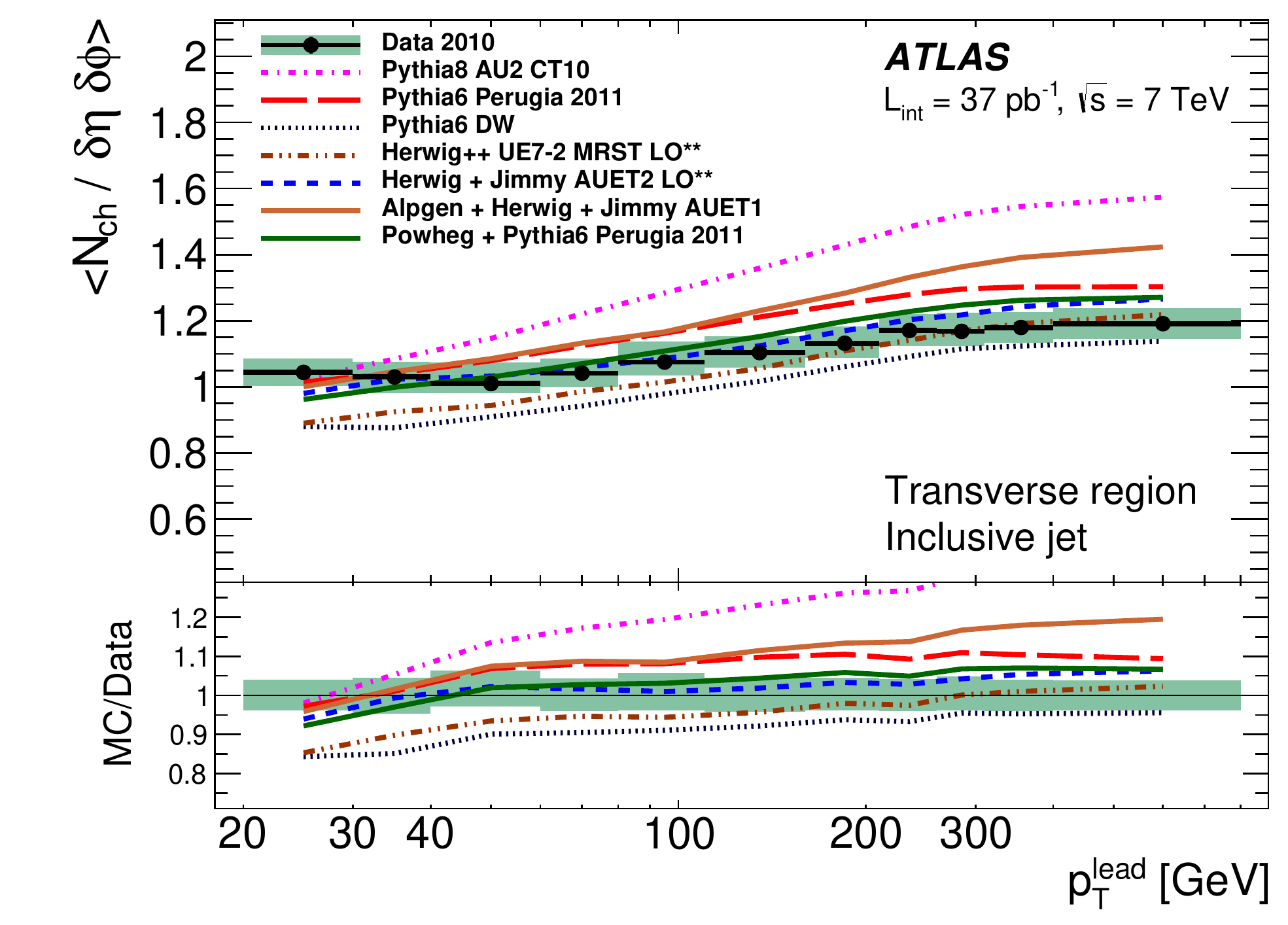}{fig:in-nchg}
    \resultsimg{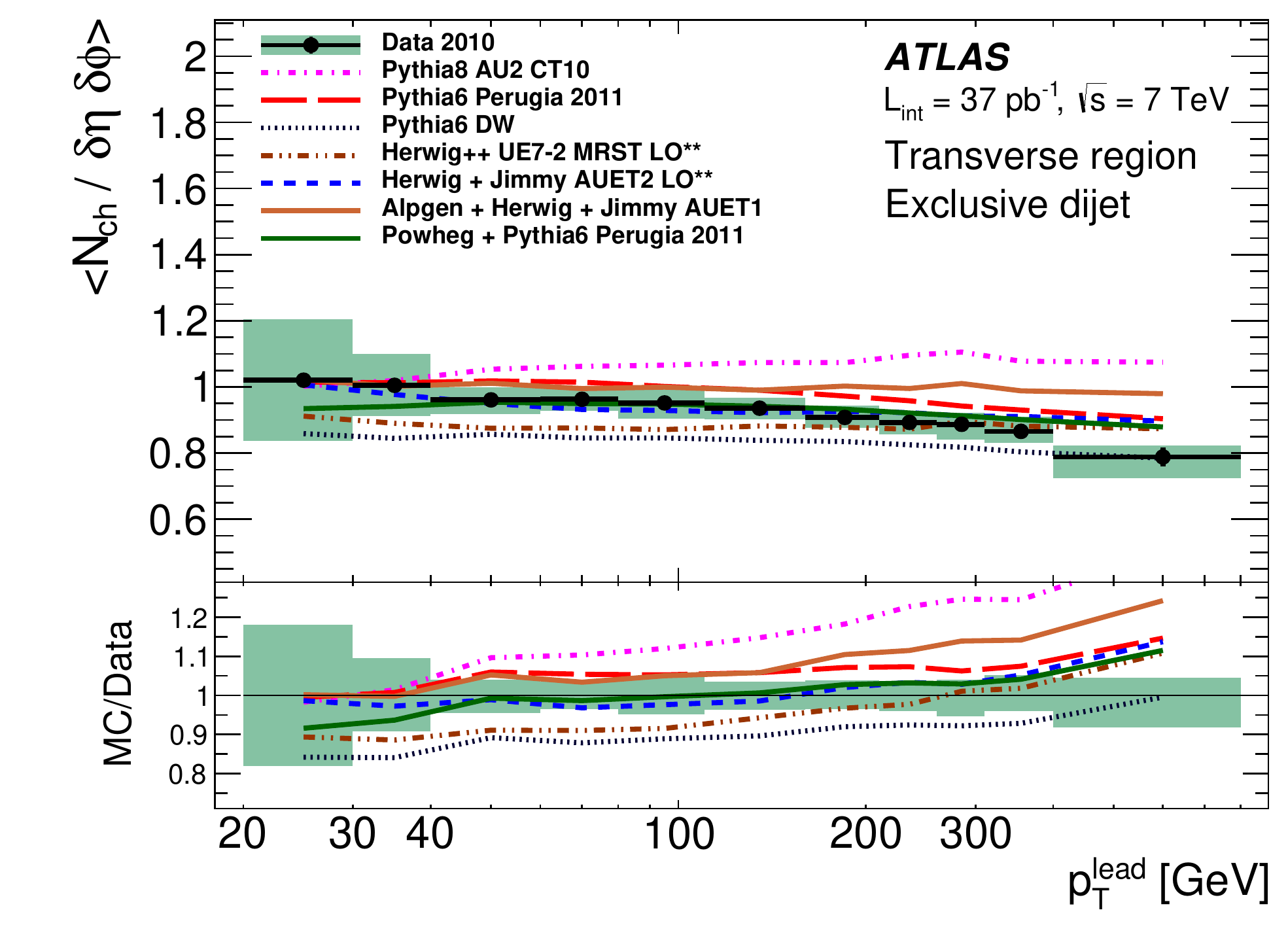}{fig:ex-nchg}
  \end{center}

  \caption{Profiles of charged-particle \protect\ptsum (top row) and charged
    multiplicity (bottom row) densities against the leading-jet \protect\pT, for the
    inclusive jet (left column) and exclusive dijet (right column) event selection.
    The jets are required to have \protect\pT of at least $20\;\GeV$, and be
    within $|y| < 2.8$, whereas the charged particles have at least a \protect\pT of $0.5\;\GeV$
    and  $|\eta| < 2.5$.
    The total transverse-region activity is compared with several MC models, with the data
    error bars indicating the statistical uncertainty and the shaded area
    showing the combined statistical and systematic uncertainty.}
  \label{fig:trk-pro}
\end{figure*}

\begin{figure*}[tbp]
  \vspace{0.5cm}
  \begin{center}
    \resultsimg{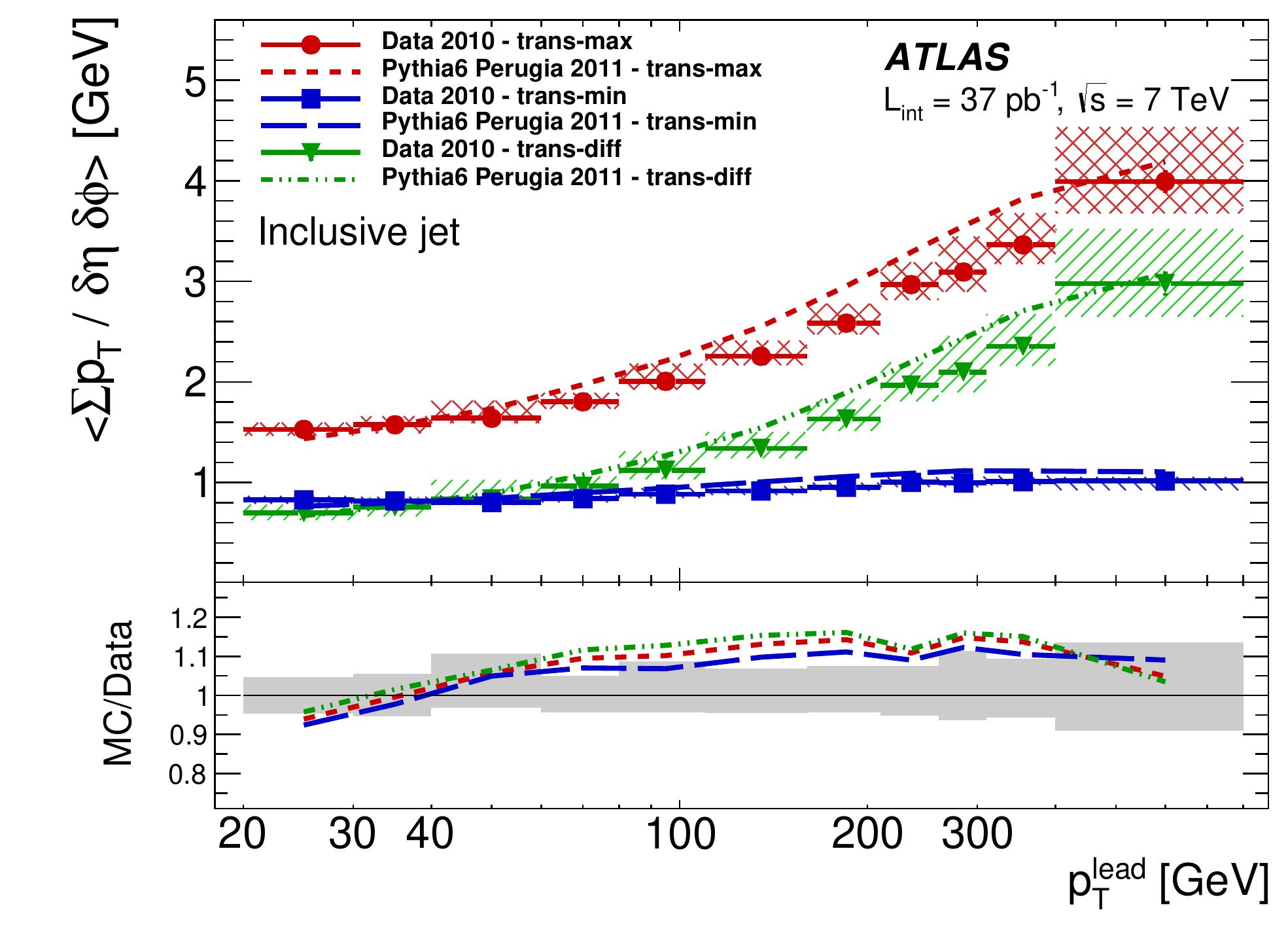}{fig:in-maxmin-ptsum}
    \resultsimg{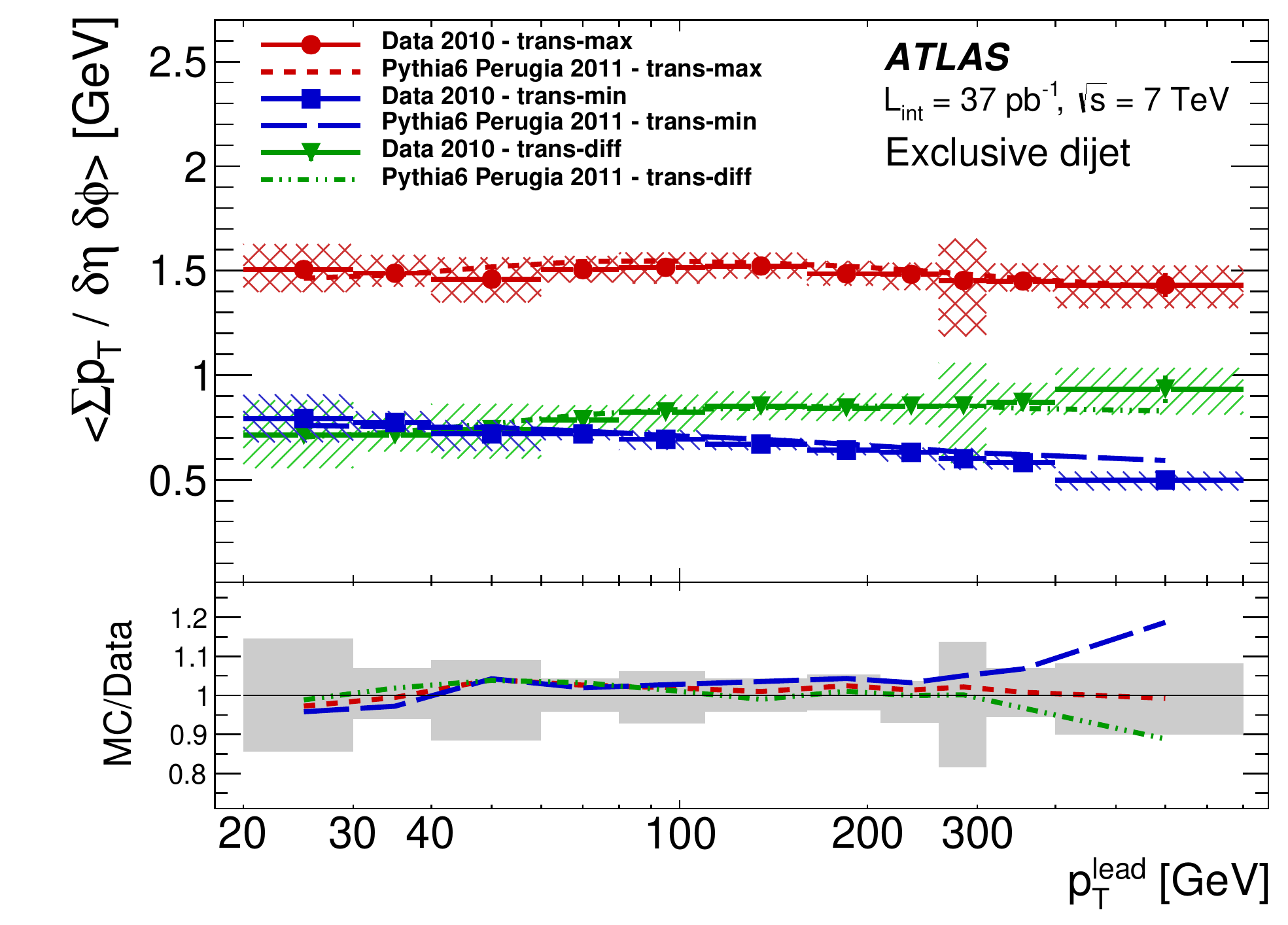}{fig:ex-maxmin-ptsum}\\
    \resultsimg{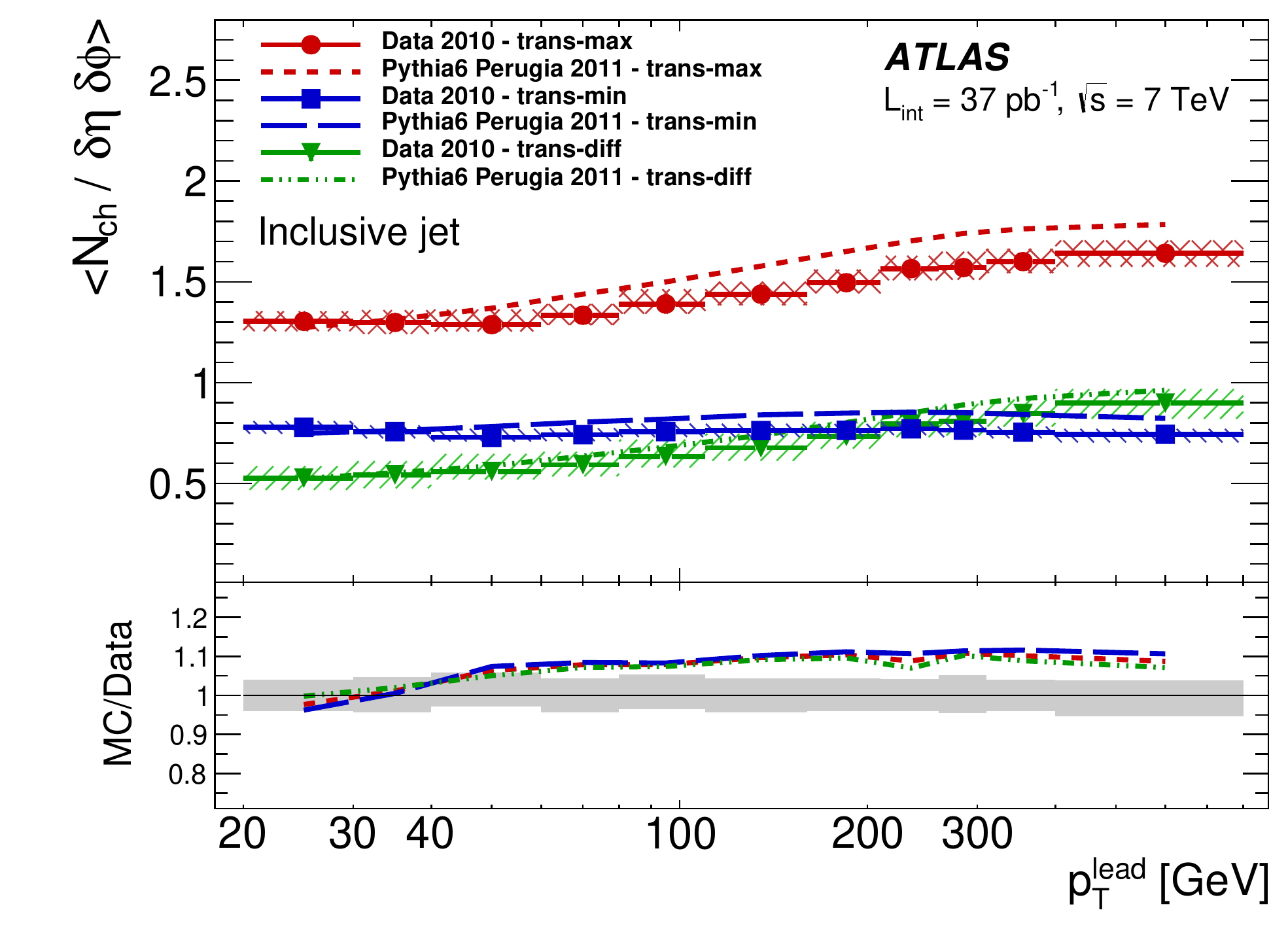}{fig:in-maxmin-nchg}
    \resultsimg{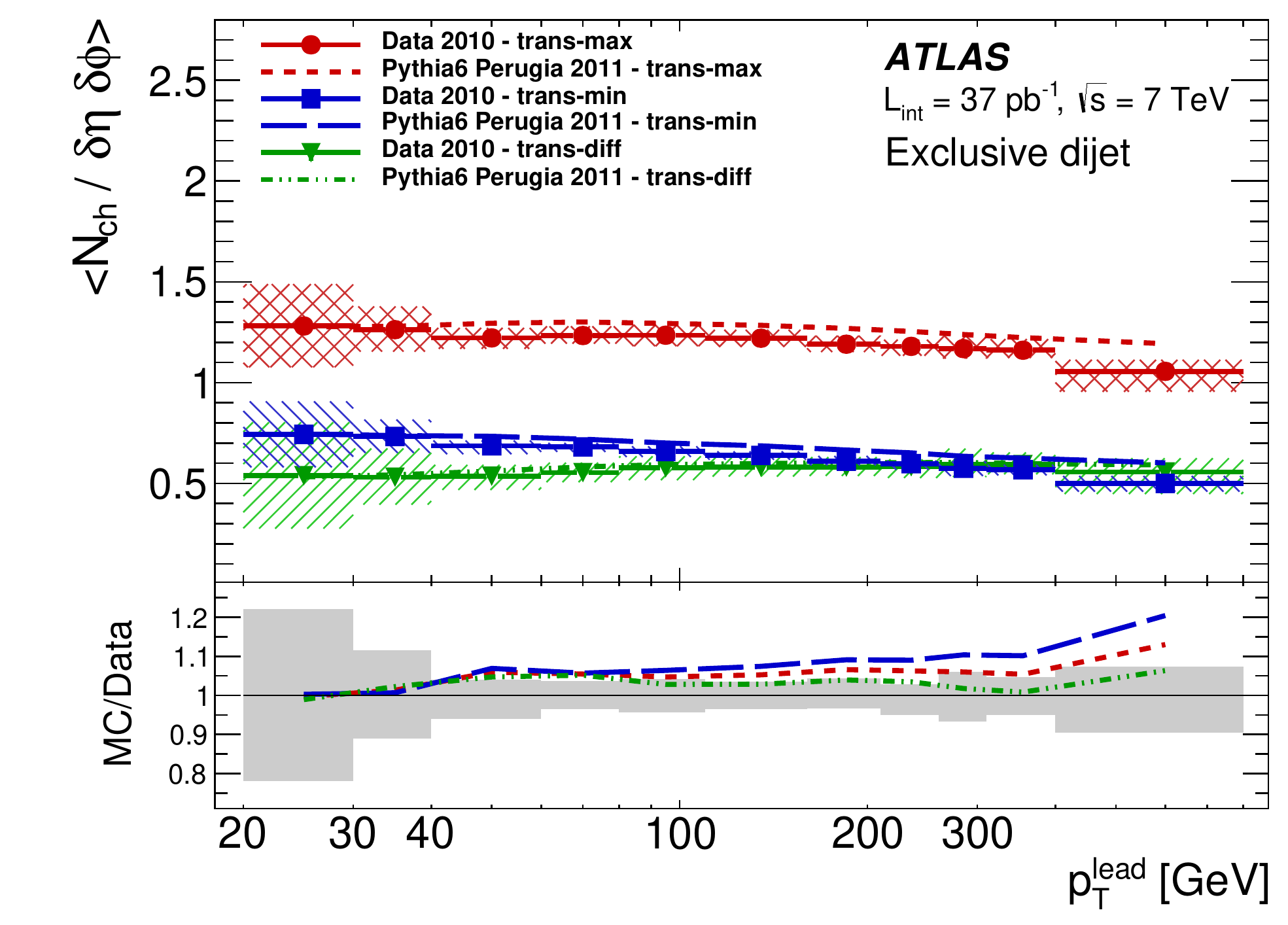}{fig:ex-maxmin-nchg}
  \end{center}

  \caption{Profiles of charged-particle \protect\ptsum (top row) and charged
    multiplicity (bottom row) densities against the leading-jet \protect\pT, for the
    inclusive jet (left column) and exclusive dijet (right column) event selection.
    The jets are required to have \protect\pT of at least $20\;\GeV$, and
    be within $|y| < 2.8$, whereas the charged particles have at least a \protect\pT of $0.5\;\GeV$
    and  $|\eta| < 2.5$.
   The plots compare the trans-max/min/diff observables to each other and
    the \pythiasix~Perugia\,2011 MC model.
    The error bands in the top plot show the combined statistical and systematic uncertainty, while
    the grey error band in the bottom plot is the envelope of the maximum
    combined statistical and systematic uncertainty of the trans-max and trans-min regions.}
  \label{fig:trk-ol}
\end{figure*}

\begin{figure*}[tbp]
  \vspace{0.5cm}
  \begin{center}
    \resultsimg{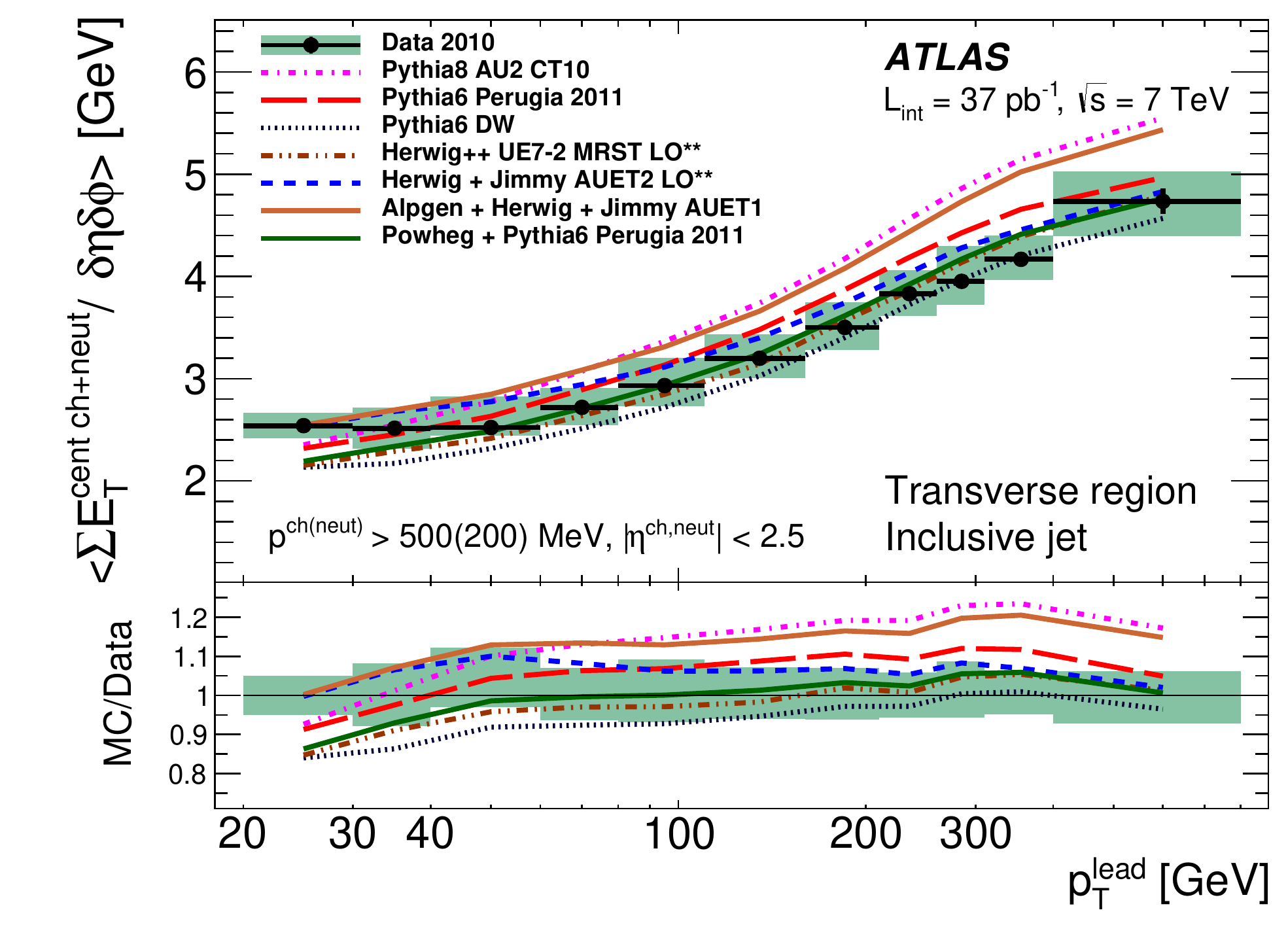}{fig:in-etsum}
    \resultsimg{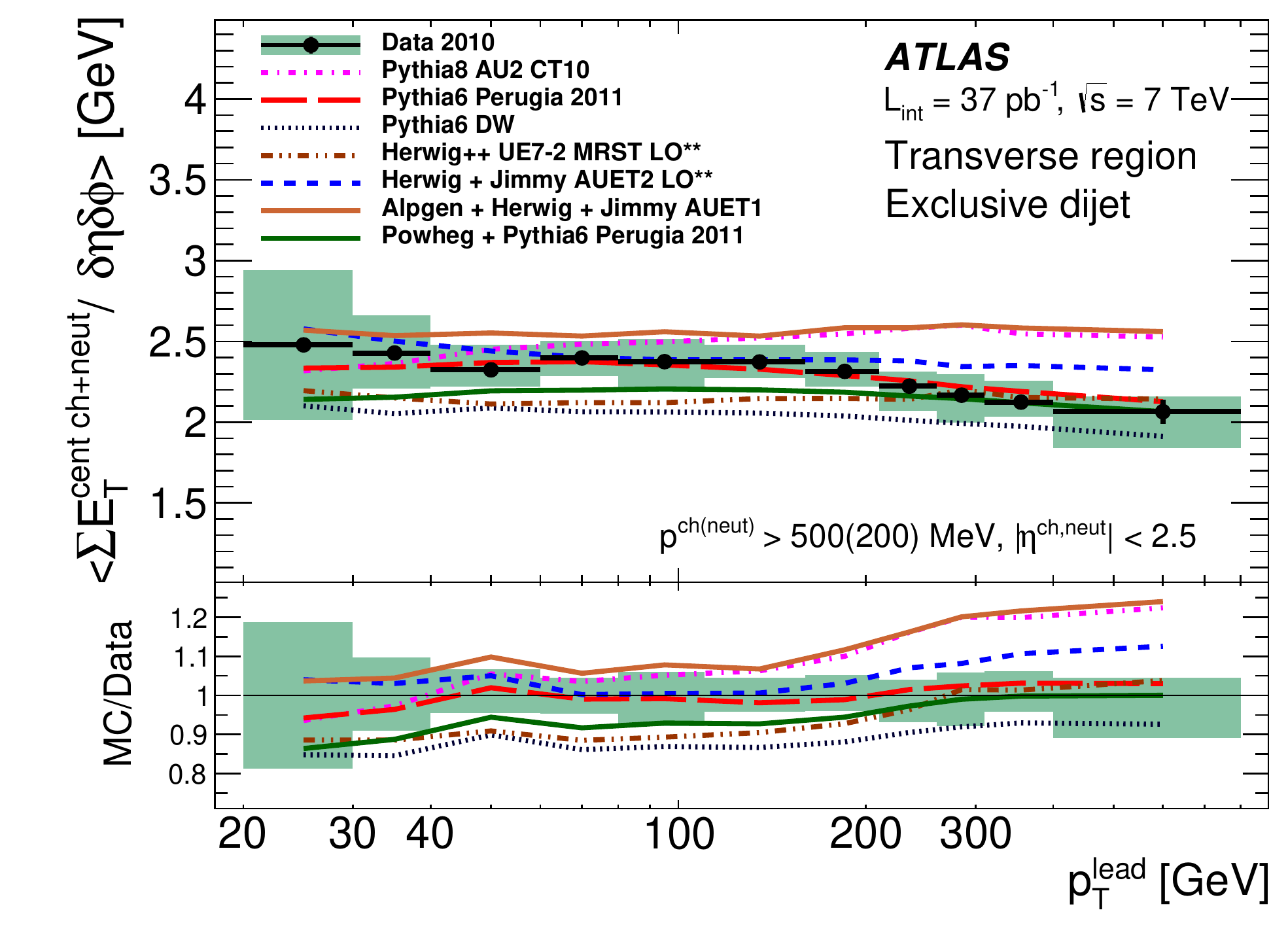}{fig:ex-etsum}\\
    \resultsimg{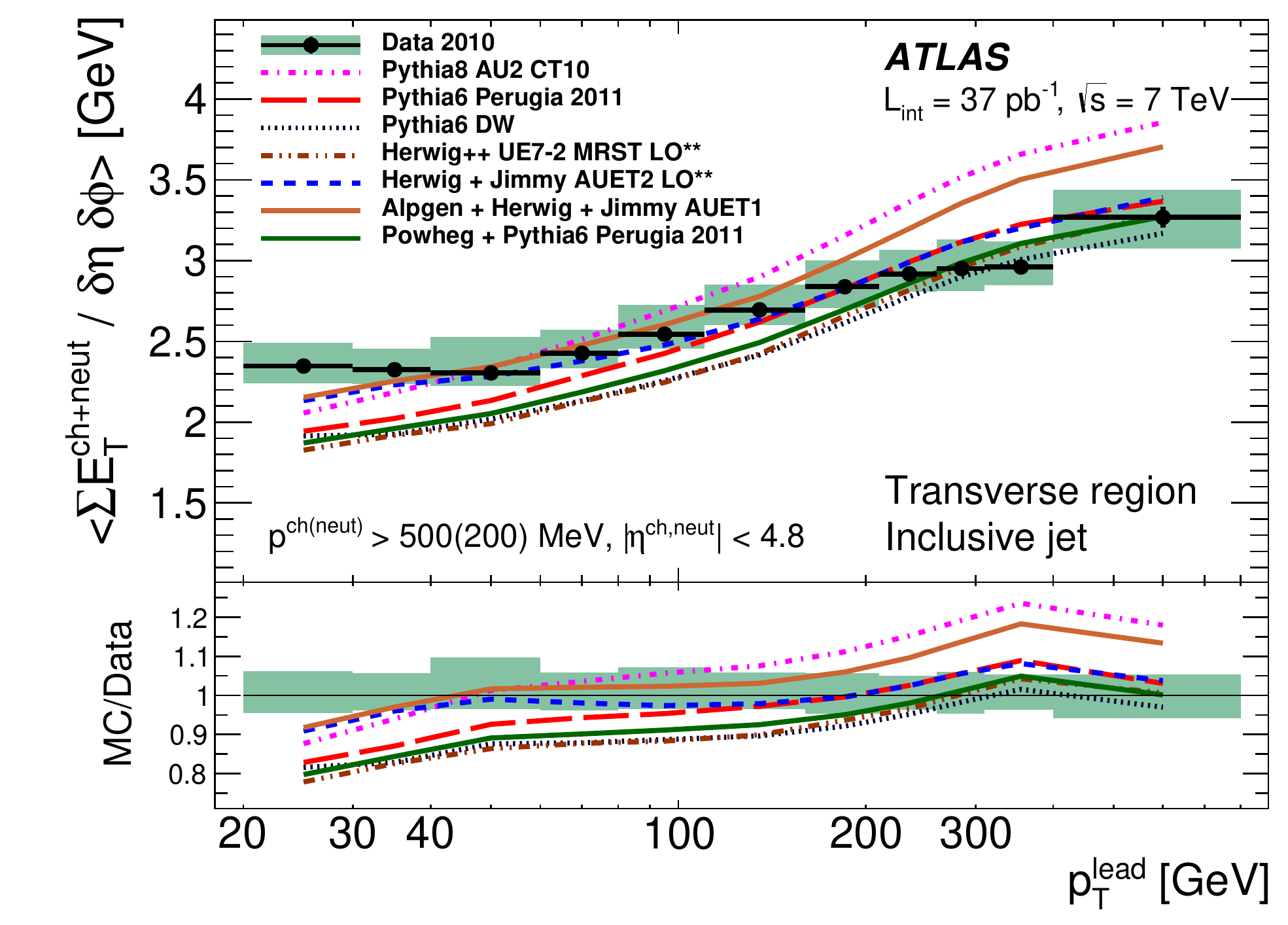}{fig:in-etsum-c}
    \resultsimg{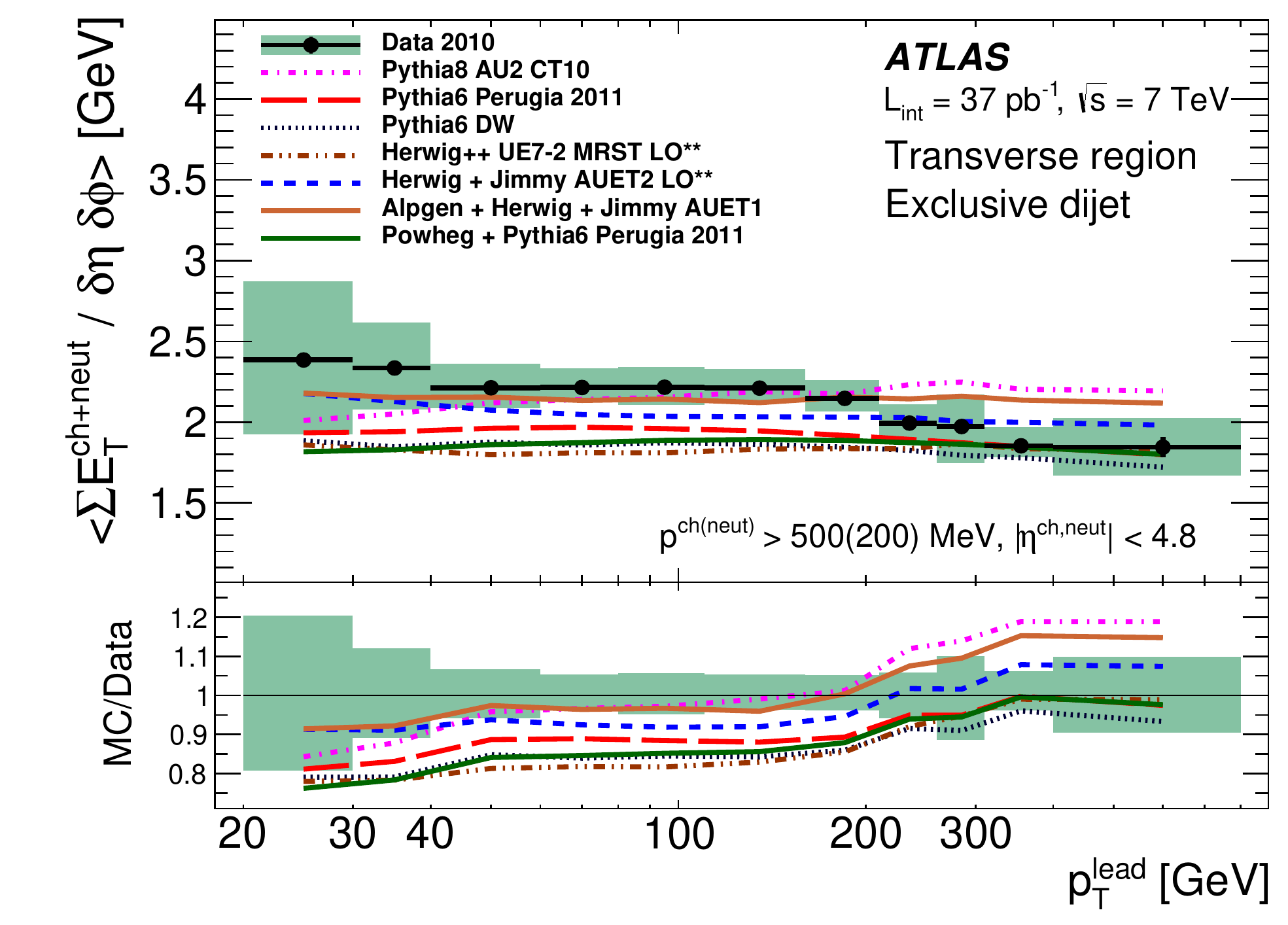}{fig:ex-etsum-c} \\
    \resultsimg{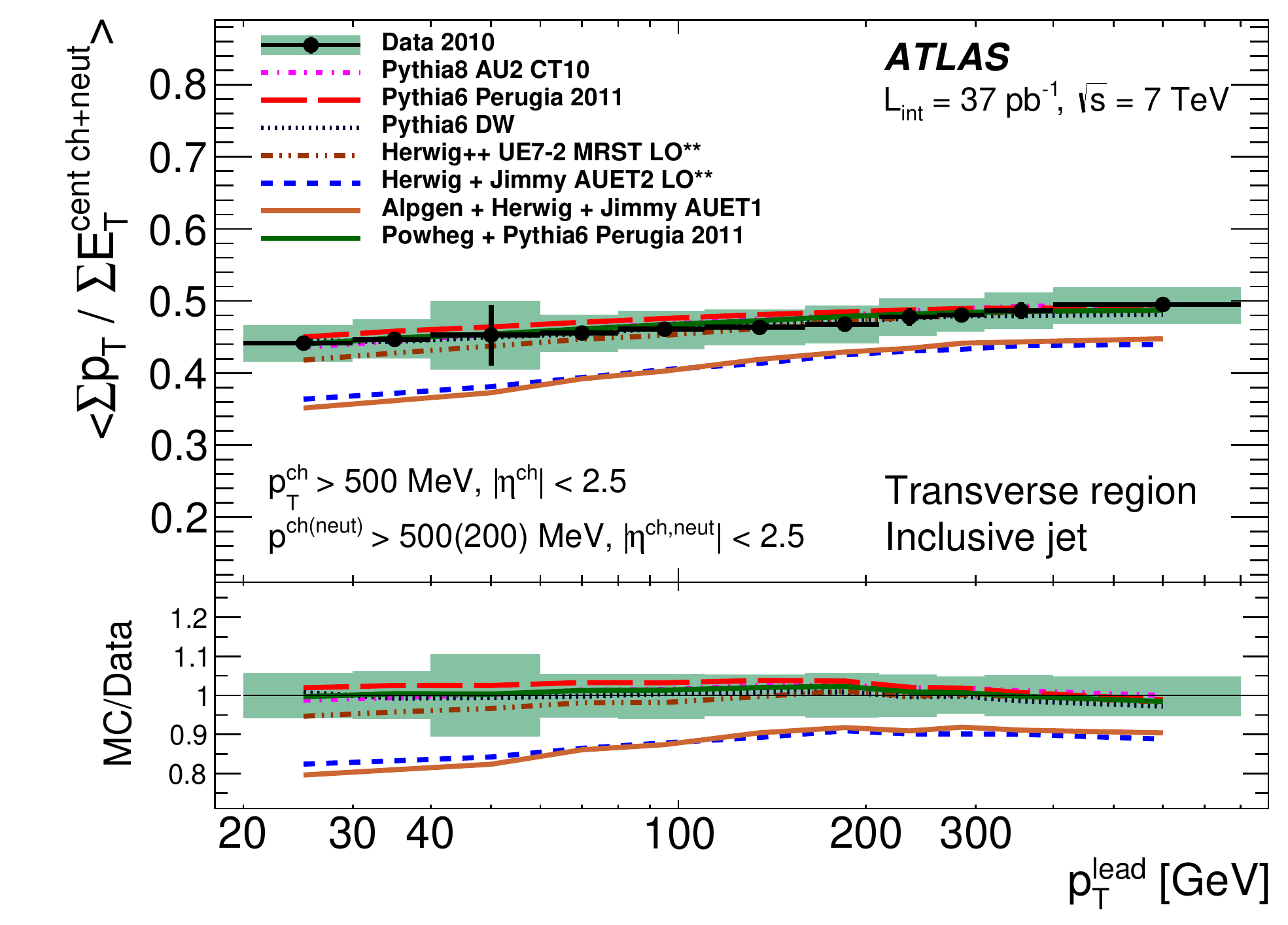}{fig:in-chgfrac}
    \resultsimg{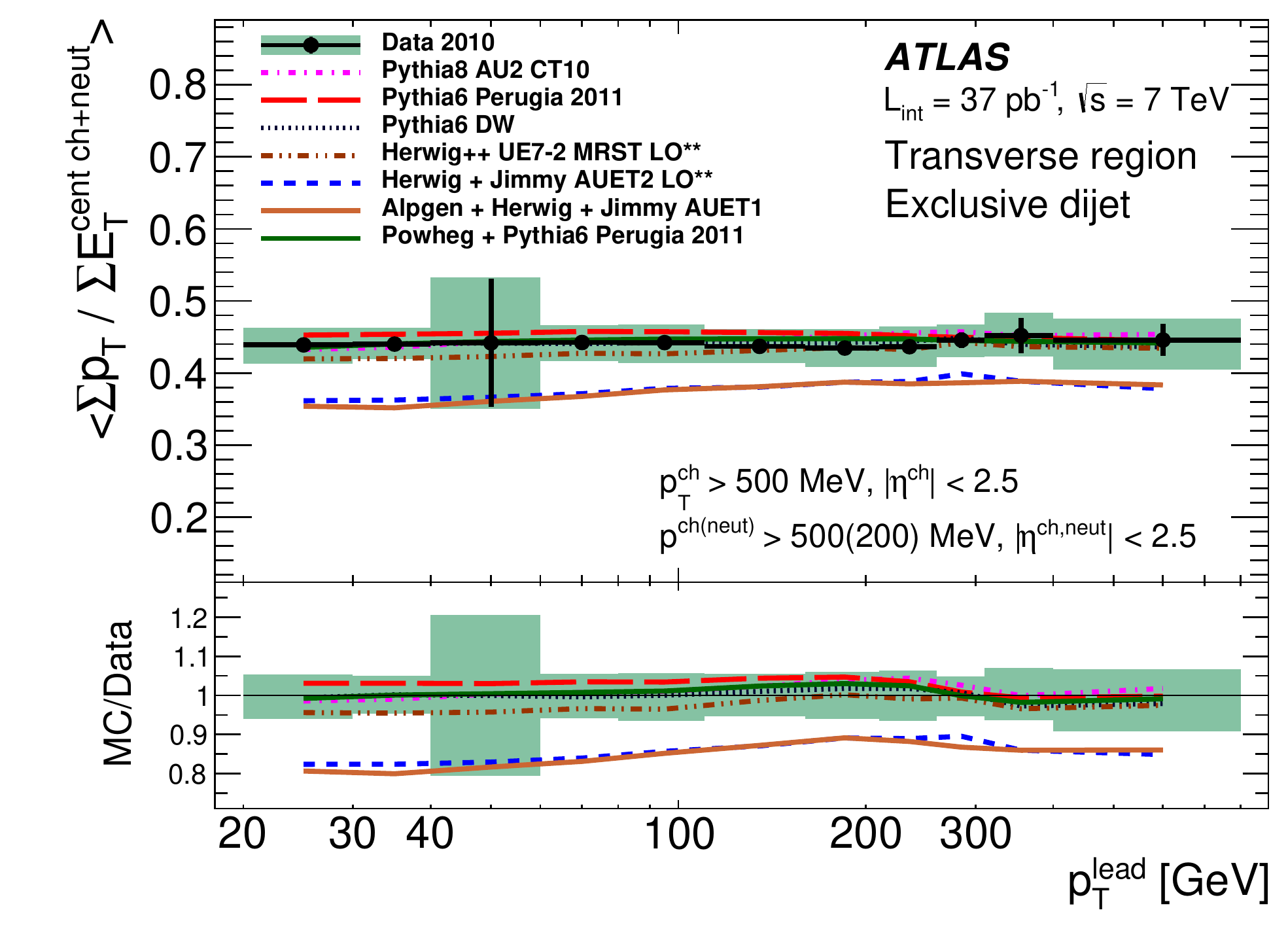}{fig:ex-chgfrac}\\
  \end{center}

  \caption{Transverse region profiles of \etsum of the neutral and charged particles
    for the inclusive (left) and exclusive dijet (right) selection against the
    leading-jet \protect\pT for the central region (top row) and the full
    acceptance region (middle row). The bottom row shows
    transverse region profiles of $\langle \ptsum(\text{charged}) / \etsum(\text{all}) \rangle$ for the
    inclusive (left) and exclusive dijet (right) selection constructed from charged and neutral particles
    against the leading-jet \protect\pT.
    The jets are required to have \protect\pT of at least $20\;\GeV$, and
    be within $|y| < 2.8$.
    The error bars show the statistical
    uncertainty while the shaded area shows the combined statistical and
    systematic uncertainty.}
  \label{fig:pro-clus}
\end{figure*}

\begin{figure*}[tbp]
  \vspace{0.5cm}
  \begin{center}
    \resultsimg{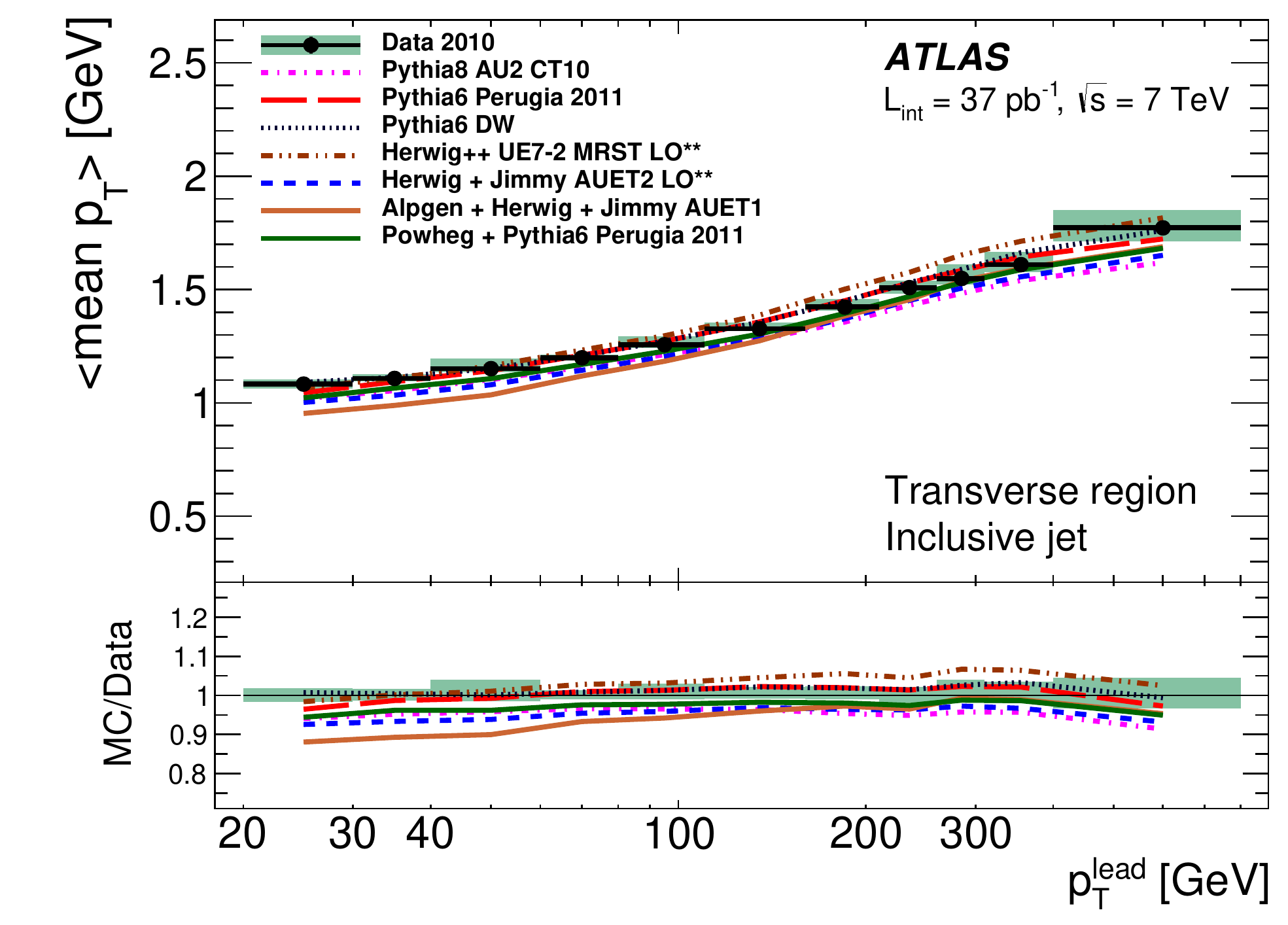}{fig:in-meanpt}
    \resultsimg{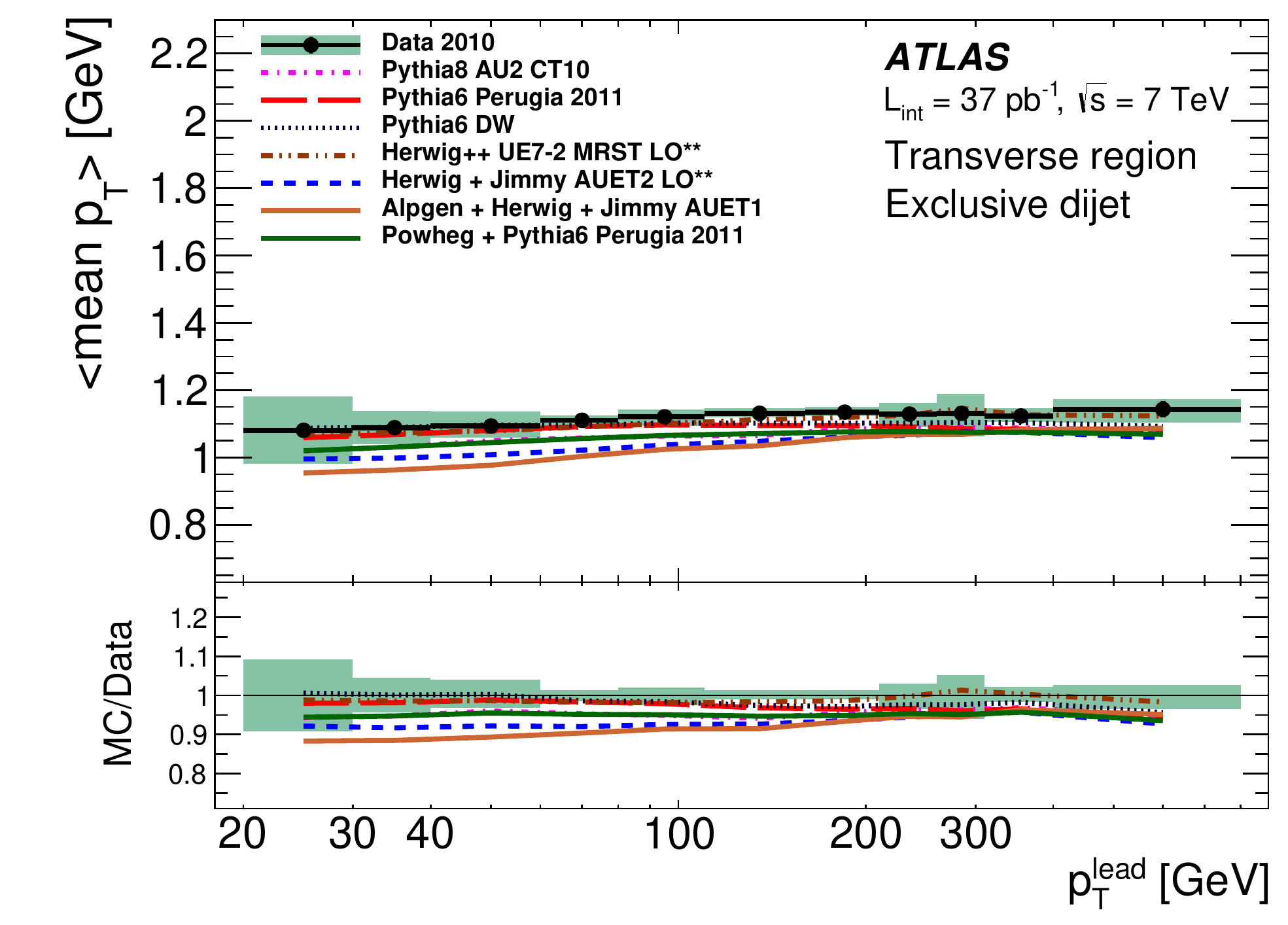}{fig:ex-meanpt}\\
    \resultsimg{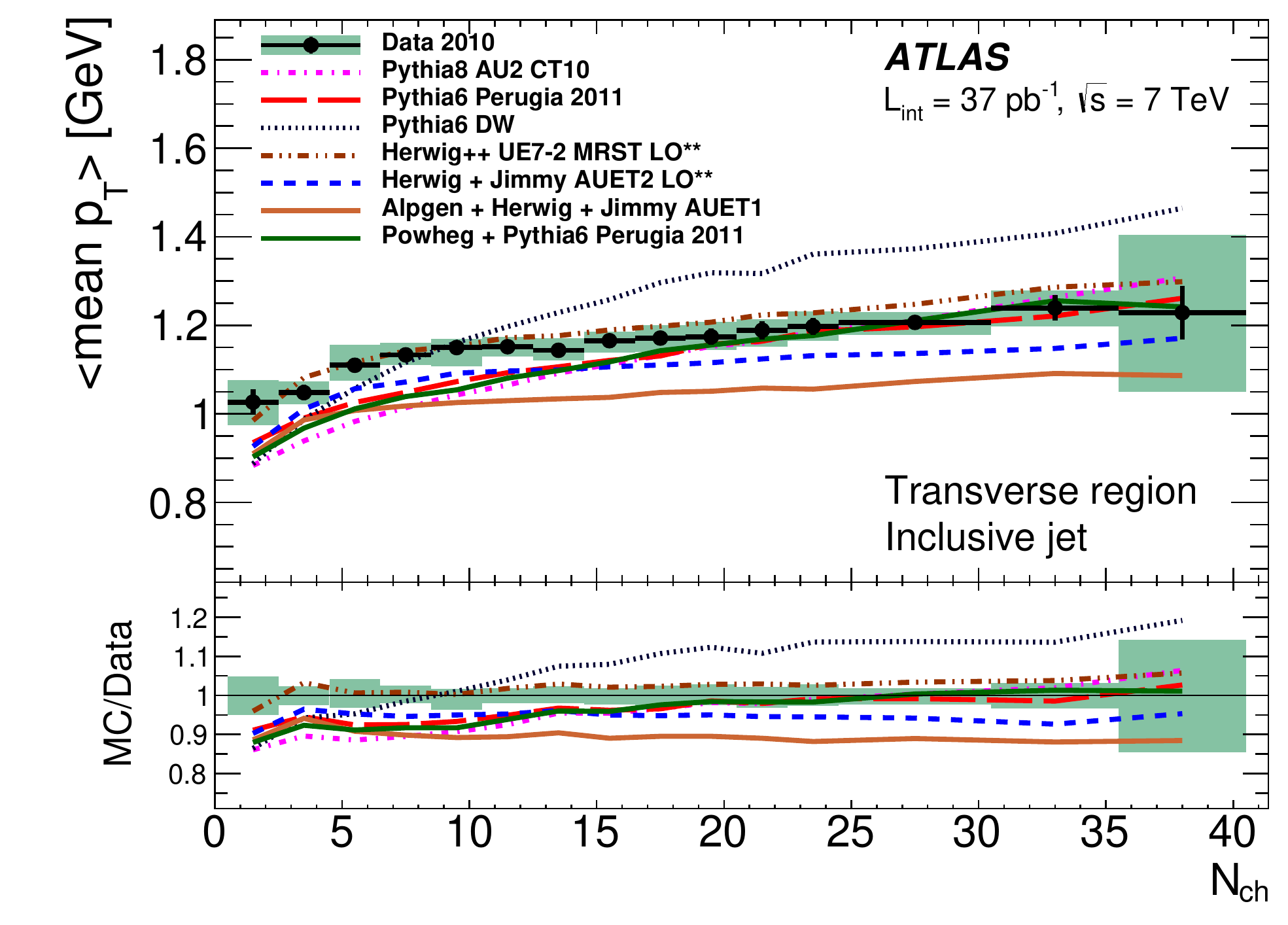}{fig:in-ptcorr}
    \resultsimg{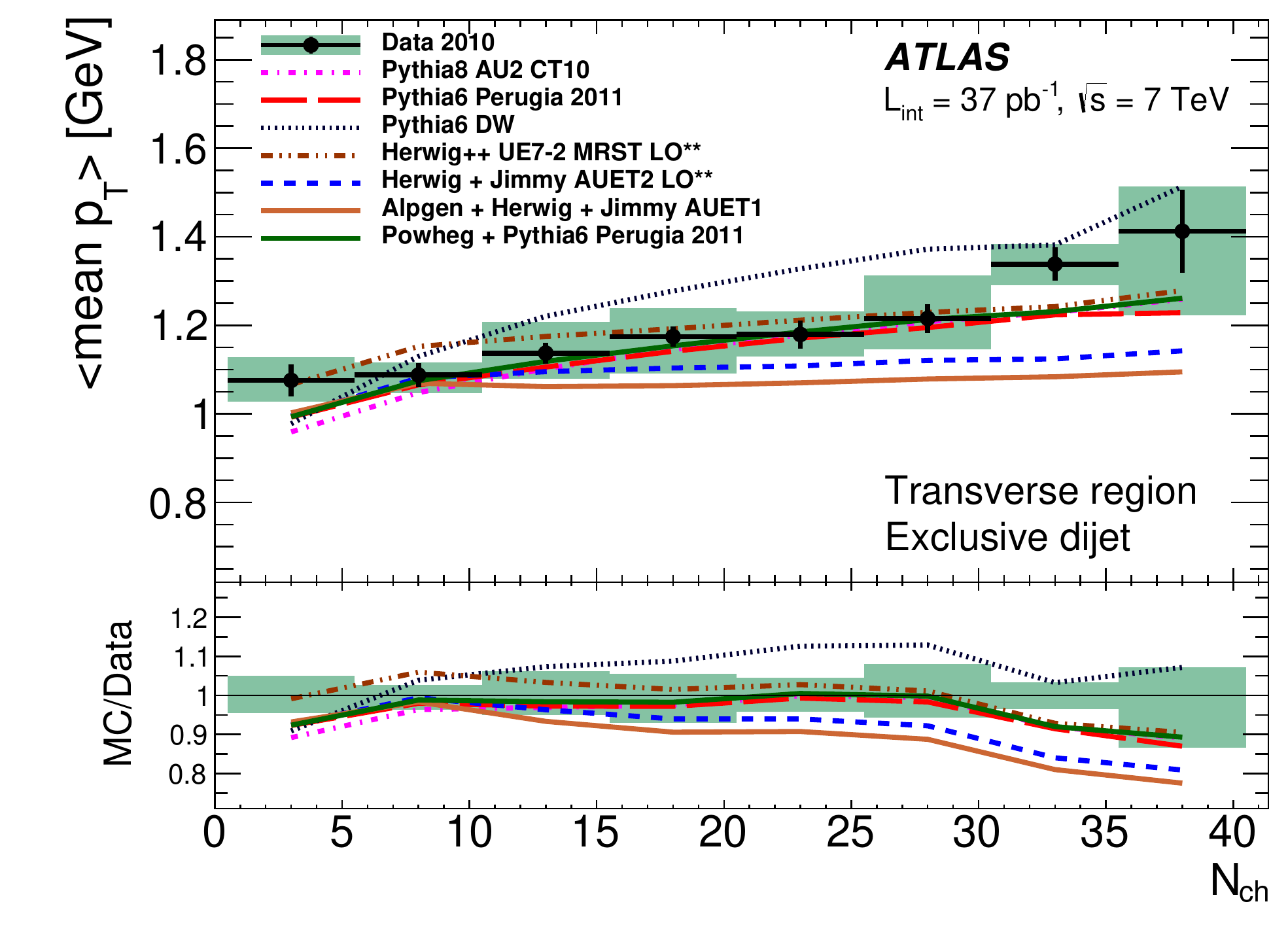}{fig:ex-ptcorr}
  \end{center}

  \caption{Transverse region profiles of the mean \protect\pT of charged particles
    for the inclusive (left) and exclusive dijet (right) selection against the
    leading-jet \protect\pT (top row) and charged-particle multiplicity (bottom row).
    The jets are required to have \protect\pT of at least $20\;\GeV$ and to be
    within $|y| < 2.8$, whereas the charged particles have at least a \protect\pT of $0.5\;\GeV$
    and  $|\eta| < 2.5$.
    The error bars show the statistical uncertainty while the shaded area shows
    the combined statistical and systematic uncertainty.}
  \label{fig:pro-meanpt}
\end{figure*}


\section{Results}
\label{sec:res}

This section presents and discusses the key distributions from this
analysis, primarily profile observables of mean UE characteristics as
functions of \ptlead (i.e. hard process scale), but also distributions of
charged-particle \ptsum and multiplicity density within bins of \ptlead, and the correlation of
transverse-region mean \pT with the charged-particle multiplicity.

In these plots, all observables are studied for both the inclusive jet
and exclusive dijet event selections. The data, corrected to particle level, are
compared to predictions of \pythiasix with the Perugia\,2011 and DW tunes,
\herwigjimmy with the AUET2 tune, \pythiaeight with the AU2~CT10 tune, \herwigpp
with the UE7-2 tune, \alpgenherwigjimmy with the AUET1 tune, and
\powhegpythia with the Perugia\,2011 tune.

To allow direct comparison between full transverse and trans-min/max/diff region
quantities, between central and full-$\eta$ cluster quantities, and to other
experiments with different angular acceptances, the observables are presented as
densities in \etaphi space. Specifically, the raw quantities are divided
by the total angular area of the region in which they were measured to produce
the number, transverse momentum, and transverse energy densities:
\Nchgdetadphi, \pTsumdetadphi, and \ETsumdetadphi respectively. The central
transverse-region observables are scaled by $\detadphi = (2 \times 2.5) \times
(2 \times \pi/3) = 10\pi/3$.  The total transverse energy of charged and neutral
particles, \etsum, is defined using the full pseudorapidity range in the
transverse region, so its area normalisation is $\detadphi = (2 \times 4.8)
\times (2 \times \pi/3) = 19.2\pi/3$.  The trans-max, trans-min and trans-diff
regions have only half of the corresponding
transverse-region area since by definition they only consider one side in $\dphi$.\\

\subsection{Distributions of \nch and \ptsum densities}
\label{sec:1d}

In \FigRef{fig:nchgsumpt-dataonly} the distributions of \nch and \ptsum densities in the
transverse, trans-max and trans-min regions, for the inclusive jet selection,
are shown for
\ptlead bins of 20--60\;\GeV, 60--210\;\GeV, and $>210\;\GeV$. This presentation
displays the evolution of each distribution in bins of increasing \ptlead, where
each \ptlead bin is independently normalised to unity.  For both the
\ptsum and \nch density observables, particularly in the transverse and trans-max
regions shown in the top two rows, the development of hard tails in the
distributions is clearly seen as the \ptlead cut is increased.  The trans-min
region in the bottom row also shows evolution in the tails, but the total
contribution to the normalisation from the tails is much less, as may be seen
from the lack of \ptlead dependence in the peak height for the trans-min region plots
as compared to the transverse and trans-max ones. The peak regions for both
variables are narrower and biased toward low values for trans-min, and wider for
trans-max.

These characteristics of \ptlead and region dependence, and the importance of
tails at high values, indicate that extra jet activity is responsible for the
changes in the transverse and trans-max region distributions, while the relative
stability of the trans-min peak region against jet activity reflects the
particular suitability of the trans-min region observables for measurement of
the multi-parton scattering contribution to the UE.

\subsection{Charged-particle \ptsum and multiplicity densities vs. \ptlead}
\label{sec:chg-ptnch-vs-ptlead}

In \FigRef{fig:trk-pro} the transverse region charged-particle multiplicity
and \ptsum density profiles are shown as functions of \ptlead.
For the inclusive jet events in the left column, the total transverse region
activity increases with \ptlead  according to both the mean \ptsum
and \nch density measures.
The exclusive dijet topology shown in the right column, where multi-jet events
are explicitly excluded, provides an alternative view of the same observables.
The exclusive dijet \ptlead profiles are seen to decrease as \ptlead increases,
although the dependence is much weaker than the opposite behaviour seen in the
inclusive jet events. This behaviour is somewhat surprising and is not
understood in detail, however we note that in addition to excluding events with
extra jets from the hard process, the exclusive dijet requirement also excludes
events with extra jets produced by MPI. This unavoidable consequence of the
dijet cut may be responsible for the downward trend which is also seen in some
MC predictions, particularly those from \pythiasix.
The data are hence broadly consistent with modelling of the UE as independent of
the hard process scale at the leading-jet \pt{} scales considered here. However
the details of the data behaviour, in particular the decreasing transverse
region activities with \ptlead in the exclusive dijet event selection are not
fully understood.

The division of the transverse regions into per-event trans-max and trans-min sides, and the
corresponding per-event differences between them, the ``trans-diff'' observables, are
shown in \FigRef{fig:trk-ol}.
In the inclusive jet events, the
trans-max activity (for both \ptsum and \nch) grows with \ptlead, similarly
to the full transverse-region trend, but its trans-min complement is almost
constant over the whole range of \ptlead.
This observation
is compatible with the interpretation of the trans-min region as being less affected
by the hard part of the underlying event.

In the max/min characterisation of exclusive dijet events, the distinction
between the behaviours of the min and max sides is reduced relative to the
inclusive jet selection, with either both slightly falling (\nch)
or the trans-max observable remaining roughly constant while the trans-min one falls (\ptsum).
As in the general transverse-region observables of \FigRef{fig:trk-pro}, this fall-off
behaviour is not fully understood, but is consistent with the exclusive dijet cuts excluding events with hard
MPI activity, an effect which is expected to be more pronounced in the trans-min
region where contamination from hard process radiation is minimised.

The insensitivity of the inclusive jet trans-min region to changes in \ptlead
indicates that UE activity can indeed be modelled as approaching a constant as a
function of hard process scale once the scale is hard enough that the proton
impact parameters are effectively zero and all collisions are central.  However,
the observed dependence of the exclusive dijet trans-min observables to changes in hard process
scale is clearly worthy of further investigation.

All MC models considered reproduce the qualitative features of the
inclusive jet data. However, the \pythia models, which have received the most UE
tuning attention in recent years, are unexpectedly further from the data than
the less flexible and less tuned \herwigpp and \herwigjimmy models on these
observables.

The spread of MC models around the exclusive dijet data is in fact slightly less than for the
inclusive jet ones, although the MC models tend not to predict quite as steep a decrease in
UE activity with \ptlead as in the data. The \pythiasix\;Perugia\,2011 model provides a good combined
description of transverse-region \ptsum and \nch densities, but \pythiaeight produces too much
transverse-region activity and \pythiasix\,DW and \herwigpp undershoot the data. A
substantial difference is seen between \herwigjimmy and the combination of
\herwig and \jimmy with \alpgen matrix elements; the \alpgenherwigjimmy
combination is more active in all observables, for both inclusive jet and
exclusive dijet selections. It is surprising that this is the case for the exclusive
dijet selection, since that is dominated by the $2 \to 2$ QCD hard process contained
in both \herwig and \alpgen.

The \powhegpythia model is seen to produce slightly lower mean UE activity than
standalone \pythiasix\;Perugia\,2011 in both the \nch and \ptsum density profile
observables and for both inclusive jet and exclusive dijet event selections,
although the two models produce similar shapes; neither is consistently closer to
data than the other, however, and a specialised shower generator tune for use
with \powheg could achieve a better data description.

Finally, it is interesting to compare the trans-diff observables between the
inclusive and exclusive jet selections; trans-diff is intended to be most
sensitive to additional hard scattering from either MPI or ISR from hard-process, and
indeed it may be seen in \FigRef{fig:trk-ol}
to be much flatter for the exclusive dijet
topology, as compared to its large increase with \ptlead for the inclusive
distributions.
Again, this behaviour is well-modelled by the
\pythiasix MC generator, with particularly good numerical agreement for the
simulation of trans-diff in the exclusive dijet selection.

\subsection{Charged and neutral particle \etsum vs. \ptlead}
\label{sec:all-et-vs-ptlead}

In \FigRef{fig:pro-clus} the corrected charged and neutral particle \etsum density is shown for both
the central region
(in the top row)
and the full
$\eta$ acceptance range (the middle row),
 for the inclusive jet and exclusive dijet topologies. The trends
are broadly similar to those for the track-based observables, and for the
central $|\eta|$ range the comparison between the data and MC models is
comparable to that seen for the equivalent charged-particle \ptsum density
plots. However, the full acceptance plots show increased disagreement between
the MC models and the data; the MC models undershoot the observed level of
activity at low \ptlead values in both the inclusive and exclusive event
selections.  This discrepancy is notable since all MPI models have to date been
tuned to observables measured solely for central rapidities. In the
full-acceptance inclusive jet observable, all models except pure \herwigjimmy
predict a faster rise of \etsum as a function of \ptlead than seen in the data,
although notably most start significantly below the data at low \ptlead.

Finally, the profiles of the ratio $\ptsum(\text{charged}) / \etsum(\text{all})$
in $|\eta| < 2.5$ against \ptlead are shown in the bottom row of
\FigRef{fig:pro-clus}. These observables are mostly flat, and are described well
by models other than those using \herwigjimmy parton showering and
hadronisation, including the \alpgen sample. These models lie 10--20\% below the
data, indicating a too-low charged fraction resulting from the \herwigjimmy
modelling. The ratio nature of this observable means that a large statistical
error is seen in the bin around 50\;\GeV, corresponding to the low-statistics
tail of events from the minimum bias trigger before the transition to the jet
trigger at \ptlead = 60\;\GeV.

\subsection{Charged-particle \ptmean vs. \ptlead and \nch}
\label{sec:chg-ptmean-vs-ptleadnch}

In \FigRef{fig:pro-meanpt}, the distributions of transverse-region
charged-particle mean \pT against \ptlead and the transverse region
charged-particle multiplicity are shown. No max/min region subdivision is made
for the \ptmean observables since while there is a clear case for identifying
the more and less active sides of an event based on particle multiplicity or
momentum flow, the physics interpretation of the transverse side with the
higher/lower average particle \pT is not as clear as for the pure \ptsum and
\nch observables.

The mean \pT vs. \ptlead profile displays a very different behaviour
between the inclusive jet and exclusive dijet event selections; in the inclusive
jet case \meanptmean rises strongly with increasing \ptlead, but when the
event selection is restricted to dijet events only, the correlation disappears
to give a distribution flat within uncertainties. The roots of this
behaviour may be seen in \FigRef{fig:trk-pro}: in the
inclusive jet case the \nch profile (the denominator in construction of \ptmean)
is less sharply rising than \ptsum (the numerator) as a function of \ptlead,
while for the exclusive dijet selection there is less distinction between \ptsum
and \nch, leading to the flat ratio. Based on previous conclusions about the
nature of the contributions to the trans-max and trans-min components of the
\ptsum and \nch density vs. \ptlead profiles, this distinction between inclusive and
exclusive mean \pT behaviours implies that it is the high-\pT tails of UE particle production
(which are effectively removed by the dijet selection) that are responsible
for the slight increase in mean \pT in the inclusive jet selection.

For both the inclusive jet and exclusive dijet selections, the transverse-region mean
\pT as a function of \ptlead is well-described by the MC models -- within
10\% of the data. This is as expected, since the descriptions of the related
transverse-region charged-particle \ptsum and multiplicity densities were described to similar
levels of precision.

The \meanptmean vs. \nch plots follow the pattern established by previous
experiments, with mean particle \pT increasing as a function of the number of
charged particles. This observable is particularly well described by \herwigpp,
which is the only generator to give a good description of the observable for
inclusive jet events with fewer than 17 particles in the transverse region.  The
bulk of the MC models considered give predictions consistent with the data for
the exclusive dijet event selection. The main outliers are \pythiasix Tune~DW,
which overshoots the data in both selections for events with more than about 10
charged particles, and \herwigjimmy both standalone and in combination with
\alpgen, which significantly undershoot. The \alpgenherwigjimmy configuration, which uses
the older AUET1 tune of \jimmy, undershoots the data by the greatest
amount.

An underestimation of \meanptmean vs. \nch data are not unexpected from the \jimmy
model, as it contains no colour reconnection mechanism to redistribute
momentum in high-multiplicity events\,\cite{Buckley:2011vq}. The newer AUET2
tune of \herwigjimmy manages to do slightly better than AUET1 as used with
\alpgen, but remains significantly outside the experimental error. By comparison
\herwigpp, which has a similar MPI implementation but does use a colour
reconnection model, agrees well with the data.

Finally, the ATLAS tunes of both \pythiasix and \pythiaeight are seen to
undershoot the data for low \nch, particularly in the inclusive jet sample, but
describe the \meanptmean of higher-multiplicity events well for both event
selections. As both these tunes incorporated the equivalent of this observable
in the ATLAS leading charged-particle UE analysis\,\cite{Aad:2010fh}, the flaws
in their data description seen here are unexpected, and use of these data in
future tunes may substantially change the MPI model parameters.

\FloatBarrier


\section{Summary and conclusions}
\label{sec:summ}

ATLAS measurements sensitive to the underlying event in $37\;\invpb$ of 7\;\TeV proton--proton
collisions at the LHC have been presented,
using observables constructed with
respect to QCD jets with \pT up to 800\;\GeV. Inclusive jet and
exclusive dijet event topologies have been considered separately, and measures
of UE activity azimuthally transverse to the leading jet computed using both
charged-particle tracks and all-particle clusters of energy deposited in the
calorimeters.
The observables have been further subdivided into
trans-max and trans-min regions on an event-by-event basis depending on which
side of the event had more activity: this subdivision provides additional
discriminating power between the component processes of the UE.

The most notable features of the observables presented in this paper are as
follows:
\begin{itemize}
\item Rising levels of transverse-region activity as a function of leading-jet \pT, as
  previously observed in UE studies, are seen in the inclusive jets event
  selection, with the exception of the inclusive trans-min region.
\item Application of an exclusive dijet event selection requirement removes this
  feature, producing instead transverse-region activity measures which are constant or
  which slightly decrease with increasing leading-jet \pT.
  This decrease in activity is not obviously expected from a simple picture of
  the underlying event
  and is deserving of further investigation.
\item The behaviour of underlying event particle flow observables in exclusive
  dijet topologies indicates that pure MPI activity can largely be modelled as
  independent of hard process scale, provided that scale is hard enough that 
  all proton--proton interactions are central.
\end{itemize}

These observables have been compared to a number of MC models, using several tunes of
commonly used underlying event models. The data are broadly consistent with the
predictions of the multiple partonic scattering paradigm, although indicative of
several areas where MPI parameter tuning may improve MC data description in further
LHC studies. Again, several key features are of note:
\begin{itemize}
\item MC models in general provide a good qualitative description of the data
  behaviour, but there are some significant discrepancies. The \herwigjimmy and
  \herwigpp generators currently describe inclusive jet UE observables slightly better
  than the \pythia family, but the \pythiasix tunes perform better on UE in
  exclusive dijet topologies.
\item \herwigpp shows significant improvements over the \herwigjimmy
  configuration. However, \pythiaeight is seen to produce too much UE activity,
  and will require retuning of underlying event parameters to perform comparably
  to its \fortran predecessor.
\item For the exclusive dijet event selection, the decrease of transverse region
  activity in data is stronger than seen in the available MC models, indicating
  that there are aspects of the UE yet to be understood and incorporated into MC
  models.
\item Full $|\eta|$-range cluster observables show larger deviations from MC
  predictions than in the central region,
 indicating a region of phase space in which all MC models can
be improved.
\end{itemize}

These data provide a detailed measurement of the $pp$ underlying event in QCD
jet events, with momentum sums and charged-particle multiplicities measured both
as distributions and as profiles of their means as functions of the hard process
scale. The separate evaluations of the observables for inclusive jet
vs. exclusive dijet event selections, charged-particle vs. all-particle energy
flows, central vs. full $\eta$ acceptance, and transverse/trans-max/trans-min
categories enable more specific MC model comparisons than possible with
preceding public data. These measurements are hence expected to play a
significant role in the future development and tuning of MC models of the
underlying event.


\section*{Acknowledgements}


We thank CERN for the very successful operation of the LHC, as well as the
support staff from our institutions without whom ATLAS could not be
operated efficiently.

We acknowledge the support of ANPCyT, Argentina; YerPhI, Armenia; ARC,
Australia; BMWF and FWF, Austria; ANAS, Azerbaijan; SSTC, Belarus; CNPq and FAPESP,
Brazil; NSERC, NRC and CFI, Canada; CERN; CONICYT, Chile; CAS, MOST and NSFC,
China; COLCIENCIAS, Colombia; MSMT CR, MPO CR and VSC CR, Czech Republic;
DNRF, DNSRC and Lundbeck Foundation, Denmark; EPLANET, ERC and NSRF, European Union;
IN2P3-CNRS, CEA-DSM/IRFU, France; GNSF, Georgia; BMBF, DFG, HGF, MPG and AvH
Foundation, Germany; GSRT and NSRF, Greece; ISF, MINERVA, GIF, I-CORE and Benoziyo Center,
Israel; INFN, Italy; MEXT and JSPS, Japan; CNRST, Morocco; FOM and NWO,
Netherlands; BRF and RCN, Norway; MNiSW and NCN, Poland; GRICES and FCT, Portugal;
MNE/IFA, Romania; MES of Russia and ROSATOM, Russian Federation; JINR; MSTD,
Serbia; MSSR, Slovakia; ARRS and MIZ\v{S}, Slovenia; DST/NRF, South Africa;
MINECO, Spain; SRC and Wallenberg Foundation, Sweden; SER, SNSF and Cantons of
Bern and Geneva, Switzerland; NSC, Taiwan; TAEK, Turkey; STFC, the Royal
Society and Leverhulme Trust, United Kingdom; DOE and NSF, United States of
America.

The crucial computing support from all WLCG partners is acknowledged
gratefully, in particular from CERN and the ATLAS Tier-1 facilities at
TRIUMF (Canada), NDGF (Denmark, Norway, Sweden), CC-IN2P3 (France),
KIT/GridKA (Germany), INFN-CNAF (Italy), NL-T1 (Netherlands), PIC (Spain),
ASGC (Taiwan), RAL (UK) and BNL (USA) and in the Tier-2 facilities
worldwide.

%
\bibliographystyle{atlasnote}
\bibliography{refs}

\onecolumn
\clearpage
\input{atlas_authlist}

\end{document}

%% file: preamble.tex
\newcommand{\invpb}{\ensuremath{\text{pb}^{-1}}\xspace}

\newcommand{\mm}{\ensuremath{\text{mm}}\xspace}
\newcommand{\eV}{\ensuremath{\text{e\kern-0.15ex{}V}}\xspace}
\newcommand{\MeV}{\ensuremath{\text{M\eV}}\xspace}
\newcommand{\GeV}{\ensuremath{\text{G\eV}}\xspace}
\newcommand{\TeV}{\ensuremath{\text{T\eV}}\xspace}
\providecommand{\unit}[2]{\ensuremath{#1\;#2}\xspace}

\newcommand{\plusplus}{\raisebox{0.2ex}{\smaller++}}

\newcommand{\fortran}{\textsc{Fortran}\xspace}

\newcommand{\alpgen}{\textsc{Alpgen}\xspace}
\newcommand{\pythia}{\textsc{Pythia}\xspace}
\newcommand{\pythiasix}{\textsc{Pythia}\,6\xspace}
\newcommand{\pythiaeight}{\textsc{Pythia}\,8\xspace}
\newcommand{\jimmy}{\textsc{Jimmy}\xspace}
\newcommand{\herwig}{\textsc{Herwig}\xspace}
\newcommand{\herwigjimmy}{\textsc{Herwig}+\allowbreak\textsc{Jimmy}\xspace}
\newcommand{\alpgenherwigjimmy}{\textsc{Alpgen}+\allowbreak\textsc{Herwig}+\allowbreak\textsc{Jimmy}\xspace}

\newcommand{\herwigpp}{\mbox{\textsc{Herwig}\plusplus}\xspace}

\newcommand{\powheg}{\textsc{Powheg}\xspace}
\newcommand{\powhegpythia}{\textsc{Powheg}+\allowbreak\textsc{Pythia}\,6\xspace}

\def\qqbar{\Pquark-\APquark}

\def\lostar{\ensuremath{\text{LO\raisebox{.3ex}{$\mspace{0.5mu}*$}}}\xspace}
\def\loss{\ensuremath{\text{LO\raisebox{.3ex}{$\mspace{0.5mu}*\mspace{-1mu}*$}}}\xspace}
\def\lostst{\loss}

\def\pt{\ensuremath{p_\mathrm{T}}\xspace}
\def\pT{\pt}
\def\Et{\ensuremath{E_\mathrm{T}}\xspace}

\def\et{\Et}
\def\kt{\ensuremath{k_t}\xspace}

\def\ptlead{\ensuremath{\pt^{\text{lead}}}\xspace}

\def\ptsum{\ensuremath{\sum\pt}\xspace}
\def\etsum{\ensuremath{\sum\Et}\xspace}

\def\ptmean{\ensuremath{\text{mean}\;\pt}\xspace}
\def\meanptmean{\ensuremath{\langle\ptmean\rangle}\xspace}
\def\nch{\ensuremath{N_\text{ch}}\xspace}
\def\etamod{\ensuremath{|\eta|}\xspace}
\def\etaphi{\ensuremath{\eta\text{--}\phi}\xspace}
\def\Nch{\nch}

\newcommand{\dphi}{\ensuremath{\Delta\!\phi}\xspace}
\newcommand{\detadphi}{\ensuremath{\delta\!\eta \, \delta\!\phi}\xspace}
\newcommand{\Nchgdetadphi}{\ensuremath{\Nch/\detadphi}\xspace}
\newcommand{\pTsumdetadphi}{\ensuremath{\ptsum/\detadphi}\xspace}
\newcommand{\ETsumdetadphi}{\ensuremath{\etsum/\detadphi}\xspace}

\newcommand{\alphaS}{\ensuremath{\alpha_\textrm{S}}\xspace}

\sfcode`\.=1001
\sfcode`\?=1001
\sfcode`\!=1001
\sfcode`\:=1001

\usepackage{ifthen}
\newcommand{\EqRef}[1]{\ifthenelse{\boolean{hmode}\and\spacefactor=1001}{Equation}{eq.}~\ref{#1}\xspace}
\newcommand{\FigRef}[1]{\ifthenelse{\boolean{hmode}\and\spacefactor=1001}{Figure}{Fig.}~\ref{#1}\xspace}
\newcommand{\FigsRef}[1]{\ifthenelse{\boolean{hmode}\and\spacefactor=1001}{Figures}{Figs.}~\ref{#1}\xspace}
\newcommand{\TabRef}[1]{\ifthenelse{\boolean{hmode}\and\spacefactor=1001}{Table}{Tab.}~\ref{#1}\xspace}
\newcommand{\TabsRef}[1]{\ifthenelse{\boolean{hmode}\and\spacefactor=1001}{Tables}{Tabs.}~\ref{#1}\xspace}
\newcommand{\SecRef}[1]{\ifthenelse{\boolean{hmode}\and\spacefactor=1001}{Section}{Sect.}~\ref{#1}\xspace}
\newcommand{\SecsRef}[1]{\ifthenelse{\boolean{hmode}\and\spacefactor=1001}{Sections}{Sects.}~\ref{#1}\xspace}
\newcommand{\AppRef}[1]{\ifthenelse{\boolean{hmode}\and\spacefactor=1001}{Appendix}{App.}~\ref{#1}\xspace}
\newcommand{\AppsRef}[1]{\ifthenelse{\boolean{hmode}\and\spacefactor=1001}{Appendices}{Apps.}~\ref{#1}\xspace}
\newcommand{\Ref}[1]{Ref.~\cite{#1}\xspace}
\newcommand{\Refs}[1]{Refs.~\cite{#1}\xspace}

%% file: atlas_authlist.tex
\begin{flushleft}
{\Large The ATLAS Collaboration}

\bigskip

G.~Aad$^{\rm 84}$,
T.~Abajyan$^{\rm 21}$,
B.~Abbott$^{\rm 112}$,
J.~Abdallah$^{\rm 152}$,
S.~Abdel~Khalek$^{\rm 116}$,
O.~Abdinov$^{\rm 11}$,
R.~Aben$^{\rm 106}$,
B.~Abi$^{\rm 113}$,
M.~Abolins$^{\rm 89}$,
O.S.~AbouZeid$^{\rm 159}$
% ...
\bigskip
\\
% ...
$^{11}$ Institute of Physics, Azerbaijan Academy of Sciences, Baku, Azerbaijan\\
% ...
$^{21}$ Physikalisches Institut, University of Bonn, Bonn, Germany\\
% ...
$^{84}$ CPPM, Aix-Marseille Universit{\'e} and CNRS/IN2P3, Marseille, France\\
% ...
$^{89}$ Department of Physics and Astronomy, Michigan State University, East Lansing MI, United States of America\\
% ...
$^{106}$ Nikhef National Institute for Subatomic Physics and University of Amsterdam, Amsterdam, Netherlands\\
% ...
$^{112}$ Homer L. Dodge Department of Physics and Astronomy, University of Oklahoma, Norman OK, United States of America\\
$^{113}$ Department of Physics, Oklahoma State University, Stillwater OK, United States of America\\
% ...
$^{116}$ LAL, Universit{\'e} Paris-Sud and CNRS/IN2P3, Orsay, France\\
% ...
$^{152}$ Institute of Physics, Academia Sinica, Taipei, Taiwan\\
% ...
$^{159}$ Department of Physics, University of Toronto, Toronto ON, Canada\\
% ...
\end{flushleft}
